\documentclass[11pt,a4paper]{article}

\usepackage[top=3.5cm,bottom=5.5cm,left=2.5cm,right=2.5cm]{geometry} 

\usepackage{array} 

\usepackage{hyperref} 
\usepackage{amsmath,amssymb,amsthm,mathrsfs,textcomp,amscd,bbm} 

\usepackage{graphicx}
\DeclareGraphicsExtensions{.pdf,.png,.jpg,.mps,.ps,}
\graphicspath{{figures/}}

\usepackage{braket}

\newcommand{\be}{\begin{equation}}
\newcommand{\ee}{\end{equation}}
\newcommand{\bea}{\begin{eqnarray}}
\newcommand{\eea}{\end{eqnarray}}
\newcommand{\bi}{\begin{itemize}}
\newcommand{\ei}{\end{itemize}}

\newcommand{\bspl}{\begin{split}}
\newcommand{\espl}{\end{split}}

\newcommand{\MeV}{\,\mathrm{MeV}}

\newcommand{\GeV}{\,\mathrm{GeV}}
\newcommand{\fm}{\,\mathrm{fm}}

\newcommand{\CKM}{\mathrm{CKM}}

\def\mev{{\rm MeV}}
\def\gev{{\rm GeV}}
\def\tev{{\rm TeV}}
\def\fm{{\rm fm}}
 
\def\fm{\mathrm{fm}}
\def\ev{\mathrm{e\kern-0.1em V}}
\def\kev{\mathrm{ke\kern-0.1em V}}
\def\mev{\mathrm{Me\kern-0.1em V}}
\def\gev{\mathrm{Ge\kern-0.1em V}}
\def\tev{\mathrm{Te\kern-0.1em V}}

\begin{document}

\begin{center}

{\bf \LARGE Scalar and vector form factors of \boldmath$D \to \pi(K) \ell \nu$ decays\\[4mm] with \boldmath$N_f=2+1+1$ twisted fermions}

\vspace{1cm}

{\Large V.~Lubicz$^{(a,b)}$, L.~Riggio$^{(b)}$, G.~Salerno$^{(a,b)}$, S.~Simula$^{(b)}$, C.~Tarantino$^{(a,b)}$}

\vspace{1cm}

$^{(a)}$ Dipartimento di Matematica e Fisica, Universit\'a di Roma Tre,\\ Via della Vasca Navale 84, I-00146 Roma, Italy\\[2mm]
$^{(b)}$ Istituto Nazionale di Fisica Nucleare, Sezione di Roma Tre,\\ Via della Vasca Navale 84, I-00146 Roma, Italy

\end{center}

\begin{figure}[htb!]
\centering{\includegraphics[scale=0.2]{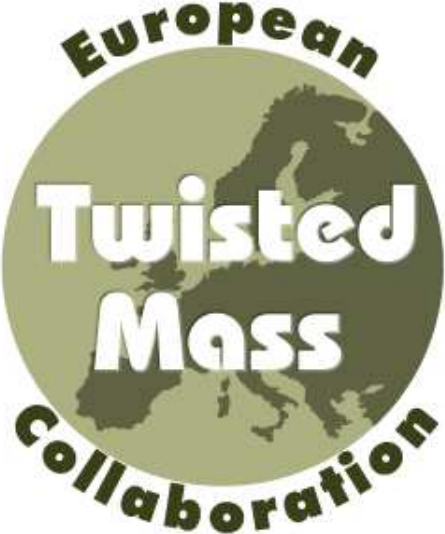}}
\end{figure}

\begin{abstract}
We present a lattice determination of the vector and scalar form factors of the $D \to \pi \ell \nu$ and $D \to K \ell \nu$ semileptonic decays, which are relevant for the extraction of the CKM matrix elements $\lvert V_{cd} \rvert$ and $\lvert V_{cs} \rvert$ from experimental data. 
Our analysis is based on the gauge configurations produced by the European Twisted Mass Collaboration with $N_f = 2 + 1 + 1$ flavors of dynamical quarks, at three different values of the lattice spacing ($a \simeq 0.062, ~ 0.082, ~ 0.089$ fm) and with pion masses as small as 210 MeV. 
Quark momenta are injected on the lattice using non-periodic boundary conditions. 
The matrix elements of both vector and scalar currents are determined for a plenty of kinematical conditions in which parent and child mesons are either moving or at rest. 
Lorentz symmetry breaking due to hypercubic effects is clearly observed in the data and included in the decomposition of the current matrix elements in terms of additional form factors. 
After the extrapolations to the physical pion mass and to the continuum limit we determine the vector and scalar form factors in the whole kinematical region from $q^2 = 0$ up to $q^2_{\rm max} = (M_D - M_{\pi(K)})^2$ accessible in the experiments, obtaining a good overall agreement with experiments, except in the region at high values of $q^2$ where some deviations are visible.
A set of synthetic data points, representing our results for $f_+^{D \pi(K)}(q^2)$ and $f_0^{D \pi(K)}(q^2)$ for several selected values of $q^2$, is provided and also the corresponding covariance matrix is available.
At zero 4-momentum transfer we get: $f_+^{D \to \pi}(0) = 0.612 ~ (35)$ and $f_+^{D \to K}(0) = 0.765 ~ (31)$. 
Using the experimental averages for $\lvert V_{cd}\rvert f_+^{D \to \pi}(0)$ and $\lvert V_{cs}\rvert f_+^{D \to K}(0)$, we extract $\lvert V_{cd}\rvert = 0.2330 ~ (137)$ and $\lvert V_{cs}\rvert = 0.945 ~ (38)$, respectively.
The second-row of the CKM matrix is found to be in agreement with unitarity within the current uncertainties:  $|V_{cd}|^2 + |V_{cs}|^2 + |V_{cb}|^2 = 0.949 ~ (78)$. 
\end{abstract}

\clearpage

\section{Introduction}
\label{sec:intro}

The Cabibbo-Kobayashi-Maskawa (CKM) matrix \cite{Cabibbo:1963yz,Kobayashi:1973fv} describes the quark flavor mixing in the electroweak sector of the Standard Model (SM). 
It represents the only source of CP violation in the quark sector and the origin of its hierarchical structure, which describes so precisely the relative strength of the flavor-changing weak quark currents, is still unexplained. 
In the SM the CKM matrix is unitary, so we have that $V_\CKM \times V^\dagger_\CKM = \mathbbm{1}$. 
This gives rise to unitarity conditions between the elements of its rows and columns, that are represented by diagonal constraints and ``unitarity triangles'', whose investigation has been the focus of much of the experimental and theoretical efforts in Flavor Physics during the recent years. 
On the one hand side inconsistencies in the CKM-picture would indicate the presence of new physics beyond the SM. 
On the other hand side, if all the precision tests of the SM performed so far are in agreement with the CKM paradigm, the absence of deviations provides stringent constraints on nonstandard phenomena and their energy scale. 
It is therefore important to determine all CKM matrix elements as precisely as possible by studying flavour-changing processes both experimentally and theoretically.

The golden modes for testing the unitarity of the second row of the CKM matrix, namely $\lvert V_{cd}\rvert^2 + \lvert V_{cs}\rvert^2 + \lvert V_{cb}\rvert^2 = 1$, are represented by the leptonic and semileptonic decays of charmed $D$ and $D_s$ mesons, which probe the $c \to d$ and $c \to s$ quark transitions, respectively. 
Combining the experimental measurements of the branching fractions of these processes with the theoretical calculations of the relevant hadronic matrix elements, i.e.~the leptonic decay constants $f_D$ and $f_{D_s}$ and the semileptonic vector form factors at zero four-momentum transfer $f_+^{D \to \pi}(0)$ and $f_+^{D \to K}(0)$, the CKM entries $\lvert V_{cd}\rvert$ and $\lvert V_{cs}\rvert$ can be determined. 
The CKM matrix element $\lvert V_{cb}\rvert$, being of the order ${\cal{O}}(10^{-2})$, is irrelevant for the second-row unitarity test at the current level of precision.
According to the $V - A$ structure of the SM weak currents, leptonic and semileptonic $D$ and $D_s$ decays should provide consistent results for the CKM elements $\lvert V_{cd}\rvert$ and $\lvert V_{cs}\rvert$.

As is well known, the theoretical calculations of hadronic matrix elements based entirely on first principles can be properly carried out by simulating the fundamental theory of the strong interaction, QCD, on a lattice. 
Thanks to the remarkable progress in algorithms and computing machines Lattice QCD (LQCD) has entered the precision era and the accuracy of numerical computations is becoming comparable to that of experiments. 
For some relevant hadronic quantities in Flavour Physics the goal of the per cent level of precision has been already achieved (see, e.g., the FLAG review \cite{Aoki:2016frl}). 

In this work we present the first $N_f = 2+1+1$ LQCD calculation of the vector and scalar form factors $f_+^{D \pi(K)}(q^2)$ and $f_0^{D \pi(K)}(q^2)$ governing the semileptonic $D \to \pi(K) \ell \nu$ decays, using the gauge configurations generated by the European Twisted Mass Collaboration (ETMC) with $N_f = 2 + 1 + 1$ dynamical quarks, which include in the sea, besides two light mass-degenerate quarks, also the strange and charm quarks with masses close to their physical values \cite{Baron:2010bv,Baron:2011sf}.
The same ensembles have already been used to determine the leptonic decay constants $f_{D}$ and $f_{D_s}$, which have provided our leptonic determinations of $\lvert V_{cd}\rvert$ and $\lvert V_{cs}\rvert$~\cite{Carrasco:2014poa}, using the experimental values for $\lvert V_{cd}\rvert f_{D}$ and $\lvert V_{cs}\rvert f_{D_s}$.

At variance with most of the existing LQCD calculations (see, e.g., Ref.~\cite{Aoki:2016frl}), that provide only the value of the vector form factor at zero 4-momentum transfer, we have evaluated both the vector and scalar form factors in the whole accessible range of values of $q^2$ in the experiments, i.e.~from $q^2 = 0$ up to $q^2_{\rm max} = (M_D - M_{\pi(K)})^2$.
In our calculations quark momenta are injected on the lattice using non-periodic boundary conditions \cite{Bedaque:2004kc,deDivitiis:2004kq} and the matrix elements of both vector and scalar currents are determined for a plenty of kinematical conditions, in which parent and child mesons are either moving or at rest.

The data coming from different kinematical conditions exhibit a remarkable breaking of Lorentz symmetry due to hypercubic effects (see Refs.~\cite{Carrasco:2015bhi,Lubicz:2016wwx} for preliminary results) for both $D\to \pi$ and $D\to K$ semileptonic form factors.
We show also evidence that the hypercubic artifacts may be largely affected by the difference between the parent and the child meson masses.
This may represent an important warning in the case of the determination of the form factors governing semileptonic $B$-meson decays.

We present the subtraction of the hypercubic artifacts and the determination of the Lorentz-invariant semileptonic vector and scalar form factors after the combined extrapolations to the physical pion mass and to the continuum limit.
A set of synthetic data points, representing our results for $f_+^{D \pi(K)}(q^2)$ and $f_0^{D \pi(K)}(q^2)$ for several selected values of $q^2$, is provided (see Tables \ref{tab:synthetic_Dpi} and \ref{tab:synthetic_DK}) and the corresponding covariance matrix is available upon request.
Our results are compared with the momentum dependence of the experimental data from BELLE \cite{Widhalm:2006wz}, BABAR \cite{Lees:2014ihu,Aubert:2007wg}, CLEO \cite{Besson:2009uv} and BESIII \cite{Ablikim:2015ixa}, and a good overall agreement is obtained. 
Some deviations are nevertheless visible at high values of $q^2$ for both $D \to \pi \ell \nu$ and $D \to K \ell \nu$ decays.
 
At zero 4-momentum transfer the results of our study are
 \be
      f_+^{D \to \pi}(0) = 0.612 ~ (35) \qquad ~ , ~ \qquad f_+^{D \to K}(0) = 0.765  ~ (31) ~ .
 \ee 
Using the updated experimental values for $\lvert V_{cd} \rvert f_+^{D \to \pi}(0) = 0.1426 (19)$ and $\lvert V_{cs} \rvert f_+^{D \to K}(0) = 0.7226 (34)$, obtained by the Heavy Flavor Averaging Group (HFAG) in Ref.~\cite{Amhis:2016xyh}, the following values of the CKM matrix elements $|V_{cd}|$ and $|V_{cs}|$ are derived:
 \bea
     |V_{cd}| & = & 0.2330 ~ (133)_{\rm lat} ~ (31)_{\rm exp} =  0.2330 ~ (137) ~ , \\
     |V_{cs}| & = & 0.945 ~ (38)_{\rm lat} ~ (4)_{\rm exp} = 0.945 ~ (38) ~ ,
     \label{eq:results_CKM}
 \eea
where the errors are from the lattice calculation and from experiment, respectively, showing that the dominant error is the theoretical one.

Including the determination of $|V_{cb}|$ from $B$-meson decays \cite{Olive:2016xmw}, the unitarity test of the second-row of the CKM matrix yields
 \be
      |V_{cd}|^2 + |V_{cs}|^2 + |V_{cb}|^2 = 0.949 ~ (78) ~ ,
      \label{eq:unitarity_2nd_row}
 \ee
which confirms unitarity at the level of several percent.

The physics described by our simulations and used throughout this work corresponds to the isospin symmetric limit of QCD, where $m_u = m_d = m_{ud}$, and the quark electric charges are neglected.
Therefore, isospin breaking and electromagnetic corrections have to be added separately in phenomenological analyses.

The paper is organized as follows. 
In Section~\ref{sec:simulations} we describe the simulation details.
In Section~\ref{sec:sec1} we present the computation of the vector and scalar form factors $f^{D \pi(K)}_+(q^2)$ and $f^{D \pi(K)}_0(q^2)$ using the matrix elements of the weak vector and scalar quark currents relevant for the $D \to \pi(K)$ transition, obtained from 2- and 3-point correlation functions. 
In Section~\ref{sec:sec2} the evidence of Lorentz symmetry breaking in the momentum dependence of the form factors is presented and discussed. 
In Section~\ref{sec:sec3} we describe the strategy adopted in order to extract the physical, Lorentz invariant, vector and scalar form factors. 
This is based on a global fit of all the data corresponding to the time and spatial components of the vector current and to the scalar current for all the lattice ensembles, studying simultaneously the dependence on $q^2$, the light-quark mass $m_\ell$ and the lattice spacing $a$, and using a phenomenological Ansatz to describe the hypercubic effects.
In Section~\ref{sec:sec4} the results for $f_+^{D \pi(K)}(q^2)$ and $f_0^{D \pi(K)}(q^2)$ from the global fit are shown and compared with experimental data. 
The extraction of the CKM matrix elements $|V_{cd}|$ and $|V_{cs}|$ is also presented together with the test the unitarity of the second row of the CKM matrix. 
Finally our  conclusions are summarized in Sec.~\ref{sec:conclusions}.

\section{Simulation details}
\label{sec:simulations}

The gauge ensembles used in this work have been generated by the ETMC with $N_f = 2 + 1 + 1$ dynamical quarks, which include in the sea, besides two light mass-degenerate quarks, also the strange and the charm quarks \cite{Baron:2010bv,Baron:2011sf}.
The ensembles are the same adopted in Refs.~\cite{Carrasco:2014cwa} to determine the up, down, strange and charm quark masses, using the experimental value of the pion decay constant $f_\pi$ to set the lattice scale\footnote{With respect to Ref.~\cite{Carrasco:2014cwa} the number of independent gauge configurations adopted for the ensemble D15.48 has been increased to $90$ to improve the statistics.}. 
They have been used also to determine the leptonic decay constants $f_K$, $f_{D}$ and $f_{D_s}$ in Ref.~\cite{Carrasco:2014poa} and the vector and scalar form factors of the semileptonic $K \to \pi \ell \nu$ decay in Ref.~\cite{Carrasco:2016kpy}.

The gauge fields are simulated using the Iwasaki gluon action \cite{Iwasaki:1985we}, while sea quarks are implemented with the Wilson Twisted Mass Action at maximal twist \cite{Frezzotti:2000nk,Frezzotti:2003xj,Frezzotti:2003ni}. 
In order to avoid the mixing of strange and charm quarks in the valence sector we have adopted the non-unitary setup described in Ref.~\cite{Frezzotti:2004wz}, in which the valence strange quarks are regularized as Osterwalder-Seiler (OS) fermions \cite{Osterwalder:1977pc}, while the valence up and down quarks have the same action of the sea. The use of different lattice regularisations for the valence and sea heavy quarks avoids completely the effects of the strange and charm mixing without modifying the renormalization pattern of operators in massless schemes and produces only a modification of discretization effects.
Moreover, since we work at maximal twist, an automatic ${\cal{O}}(a)$-improvement \cite{Frezzotti:2003ni,Frezzotti:2004wz} is guaranteed also for our non-unitary setup.

The QCD simulations have been carried out at three different values of the inverse bare lattice coupling $\beta$, to allow for a controlled extrapolation to the continuum limit, and at different lattice volumes.
For each gauge ensemble we have used a number of gauge configurations corresponding to a separation of 20 trajectories to avoid autocorrelations.  
We have simulated quark masses in the range from $\simeq 3\, m_{ud}$ to $\simeq 12\, m_{ud}$ in the light sector, from $\simeq 0.7\, m_s$ to $\simeq 1.2\, m_s$ in the strange sector, and from $\simeq 0.7\, m_c$ to $\simeq 1.2\, m_c$ in the charm sector, where $m_{ud}$, $m_s$ and $m_c$ are the physical values of the average up/down, strange and charm quark masses respectively, as determined in Ref.~\cite{Carrasco:2014cwa}.
The lattice spacings are found to be $a = \{ 0.0885\,(36), 0.0815\,(30), 0.0619\,(18)\}\, \fm$   at $\beta = \{1.90, 1.95, 2.10\}$ respectively, the lattice volume goes from $\simeq 2$ to $\simeq 3$ fm and the pion masses, extrapolated to the continuum and infinite volume limits, range from $\simeq 210$ to $ \simeq 450\,\MeV $ (see Ref.~\cite{Carrasco:2014cwa} for further details).

In our study of the semileptonic $D \to \pi \ell \nu$ and $D \to K \ell \nu$ form factors we make use of the input parameters (values of quark masses and lattice spacings) obtained from the eight branches of the analysis carried out in Ref.~\cite{Carrasco:2014cwa}. 
The various branches differ by: 
~ i) the choice of the scaling variable, which was taken to be either the Sommer parameter $r_0/a$ \cite{Sommer:1993ce} or the mass of a fictitious pseudoscalar (PS) meson made of two strange-like quarks $a M_{s^\prime s^\prime}$; 
~ ii) the fitting procedures, which were based either on Chiral Perturbation Theory (ChPT) or on a polynomial expansion in the light quark mass (for the motivations see the discussion in Section 3.1 of Ref.~\cite{Carrasco:2014cwa}); 
~ iii)  the choice between two methods, denoted as M1 and M2 which differ by ${\cal{O}}(a^2)$ effects (see, e.g., Ref.~\cite{Constantinou:2010gr}), used to determine non-perturbatively the values of the mass renormalization constant (RC) ${\cal{Z}}_m = 1 / {\cal{Z}}_P$~\cite{Carrasco:2014cwa}.
Throughout this work the results corresponding to the various branches of the analysis are combined to form our averages and errors according to Eq.~(28) of Ref.~\cite{Carrasco:2014cwa}.

The basic simulation parameters and the masses of the  $\pi$, $K$ and $D$ mesons corresponding to each ensemble used in this work are collected in Table~\ref{tab:simu&masses}.

\begin{table}[!h]
\renewcommand{\arraystretch}{1.2}	 
\begin{center}	
\scalebox{0.85}{
\begin{tabular}{||>{$\!\!$}c<{$\!\!$}|>{$\!\!$}c<{$\!\!$}|>{$\!\!$}c<{$\!\!$}|>{$\!\!$}c<{$\!\!$}|>{$\!\!$}c<{$\!\!$}|>{$\!\!$}c<{$\!\!$}|>{$\!\!$}c<{$\!\!$}|>{$\!\!$}c<{$\!\!$}|>{$\!\!$}c<{$\!\!$}|>{$\!\!$}c<{$\!\!$}|>{$\!\!$}c<{$\!\!$}||}
\hline
ensemble&$\beta$ & $V / a^4$ &$a\mu_{sea}=a\mu_\ell$ &$a\mu_s$&$a\mu_c$&$M_\pi(\MeV)$&$M_K(\MeV)$& $M_D(\MeV)$& $L(\fm)$&$M_\pi L$\\
\hline
A30.32&$1.90$ & $32^{3}\times 64$ &$0.0030$ &$\{0.0180,$&$\{0.21256,$&$275$&$569$&$2015$&$2.84$&$3.96$\\
A40.32&& & $0.0040$ &$0.0220,$ &$0.25000,$&$315$&$578$&$2018$&$$&$4.53$\\
A50.32&& & $0.0050$ &$ 0.0260\}$ &$0.29404\}$&$351$&$578$&$2018$&$$&$5.04$\\
\cline{1-1}\cline{3-4}\cline{7-11}
A40.24&& $24^{3}\times 48 $ & $0.0040$ & & &$324$&$584$&$2024$&$2.13$&$3.49$\\
A60.24&& & $0.0060$ && &$386$&$599$&$2022$&$$&$4.17$\\
A80.24&& & $0.0080$ & &  &$444$&$619$&$2037$&$$&$4.79$\\
A100.24&& & $0.0100$ & &  &$495$&$639$&$2042$&$$&$5.34$\\
\hline
B25.32&$1.95$ & $32^{3}\times 64$ &$0.0025$ & $\{0.0155,$& $\{0.18705,$&$258$&$545$&$1950$&$2.61$&$3.42$\\
B35.32&& & $0.0035$ &$  0.0190,$ &$0.22000,$&$302$&$556$&$1944$&$$&$3.99$\\
B55.32&& & $0.0055$ &$ 0.0225\}$ &$0.25875\}$&$375$&$578$&$1959$&$$&$4.96$\\
B75.32&& & $0.0075$ & & &$436$&$600$&$1965$&$$&$5.77$\\
\cline{1-1}\cline{3-4}\cline{7-11}
B85.24&& $24^{3}\times 48 $ & $0.0085$  & & &$467$&$611$&$1974$&$1.96$&$4.63$\\
\hline
D15.48&$2.10$ & $48^{3}\times 96$ &$0.0015$ & $\{0.0123,$& $\{0.14454,$ &$220$&$526$&$1928$&$2.97$&$3.31$\\
D20.48&& & $0.0020$ &$0.0150,$ &$0.17000,$&$254$&$533$&$1933$&$$&$3.83$\\
D30.48&& & $0.0030$ & $  0.0177\}$&$0.19995\}$&$308$&$547$&$1939$&$$&$4.65$\\
\hline
\end{tabular}
}
\end{center}
\renewcommand{\arraystretch}{1.0}
\caption{\it Summary of the simulated sea and valence quark bare masses, of the $\pi$, $K$ and $D$ meson masses, of the lattice size $L$ and of the product $M_\pi L$ for the various gauge ensembles used in this work. The values of $M_K$ and $M_D$ do not correspond to the simulated strange and charm bare quark masses shown in the $5^{\rm th}$ and $6^{\rm th}$ columns, but to the renormalized strange and charm masses interpolated at the physical values $m_s^{phys}(\overline{MS},2~\rm{GeV}) = 99.6 (4.3) \MeV$ and $m_c^{phys}(\overline{MS}, 2~\rm{GeV}) = 1.176 (39) \GeV$ determined in Ref.~\cite{Carrasco:2014cwa}.}
\label{tab:simu&masses}
\end{table}

\section{Lattice calculation of the vector and scalar matrix elements}
\label{sec:sec1}

The  matrix  element of the weak vector current $\widehat{V}_\mu$ between an initial $D$-meson state and a $\pi$($K$)-meson final state decomposes, as required by the Lorentz symmetry, into the two form factors $f_+(q^2)$ and $f_-(q^2)$:
\be
    \braket{\widehat{V}_\mu} \equiv \braket{P(p_P) | \widehat{V}_\mu | D(p_D)} = (p_D + p_P)_\mu\, f_+(q^2) + (p_D - p_P)_\mu\, f_-(q^2) ~ ,
    \label{eq:matrix_element_decomposition}
\ee
where $P = \pi (K)$ can be either the pion or the kaon and the 4-momentum transfer $q$ is given by $q \equiv p_D - p_P$.
The scalar form factor $f_0$ is then defined as
\be
    f_0(q^2) = f_+(q^2) + \frac{q^2}{M_D^2 - M_P^2} ~ f_-(q^2)~,
    \label{eq:scalar_form_factor}
\ee
so that the kinematic identity $f_+(0) = f_0(0)$ is satisfied by definition. 
The scalar form factor is proportional to the 4-divergence of $\braket{\widehat{V}_\mu}$ so that, thanks to the Ward-Takahashi identity (WTI), $f_0$ can be determined from the matrix element of the scalar density $\widehat{S}$ between the $D$-meson and the $\pi$($K$)-meson states:
\be
    \braket{\widehat{S}} \equiv \braket{P(p_P) | \widehat{S} | D(p_D)} = \frac{M_D^2 - M_P^2}{m_c - m_x} ~ f_0(q^2) ~ ,
    \label{eq:scalar_form_factor_2}
\ee
where $x = \ell (s)$.

From Eqs.~(\ref{eq:matrix_element_decomposition}) and (\ref{eq:scalar_form_factor_2}) it turns out that the determination of the scalar and vector form factors can be carried out by computing the matrix elements $\braket{\widehat{V}_\mu}$ and $\braket{\widehat{S}}$. 
These two quantities can be extracted from the large (Euclidean) time distance behavior of a convenient combination of 2- and 3-point correlation functions in lattice QCD, which are defined as 
\bea
   \label{eq:C2}
   C_2^{D(P)}\left(t^\prime,\,\vec{p}_{D(P)}\right)\!\!\!\! & = & \!\!\!\!\frac{1}{L^3} \sum_{\vec{x},\vec{z}} \braket{0 \lvert P_5^{D(P)}(x)\,P_5^{D(P)\dagger}(z) 
          \rvert 0}\,e^{-i\vec{p}_{D(P)}\cdot(\vec{x}-\vec{z})}\,\delta_{t^\prime,\,(t_x-t_z)}~,\\
   \label{eq:C3}
   C^{DP}_{\widehat{\Gamma}}\left(  t,\, t^\prime,\, \vec{p}_D,\, \vec{p}_P \right)\!\!\!\! & = & \!\!\!\!\frac{1}{L^6}\sum_{\vec{x},\vec{y},\vec{z}} 
           \braket{0\lvert P_5^{P}(x)\widehat{\Gamma}(y)\,P_5^{D\dagger}(z)\rvert0}\,e^{-i\vec{p}_{D}\cdot(\vec{y}-\vec{z}) + i\vec{p}_{P}\cdot(\vec{y}-\vec{x})}\,
            \delta_{t,\,(t_y-t_z)}\,\delta_{t^\prime,\,(t_x-t_z)}~,\,\,\quad
\eea
where $t^\prime$ is the time distance between the source and the sink, $t$ is the time distance between the insertion of the current $\widehat{\Gamma} = \{ \widehat{V}_\mu, ~ \widehat{S} \}$ and the source, while $P_5^D = i\,\bar{c} \gamma_5 u$ and $P_5^{\pi(K)}=i\, \bar{d} (\bar{s}) \gamma_5 u$ are the interpolating fields of the $D$ and $\pi(K)$ mesons. 
The Wilson parameters $r$ of the two valence quarks in any PS meson are always chosen to have opposite values, $r_c = r_s = r_d = -r_u$, so that the squared PS mass differs from its continuum counterpart only by terms of ${\cal{O}}(a^2 m \Lambda_{QCD})$ \cite{Frezzotti:2003ni}.

The statistical accuracy of the correlators (\ref{eq:C2}-\ref{eq:C3}) is significantly improved by using the all-to-all quark propagators evaluated with the so-called ``one-end" stochastic method \cite{McNeile:2006bz}, which includes spatial stochastic sources at a single time slice chosen randomly (see Ref.~\cite{Frezzotti:2008dr} where the degenerate case of the electromagnetic pion form factor is discussed in details).
Statistical errors on the quantities directly extracted from the correlators are always evaluated using the jackknife procedure, while cross-correlations are taken into account by the use of the eight bootstrap samplings (with ${\cal{O}}(100)$ events each) corresponding to the eight analyses of Ref.~\cite{Carrasco:2014cwa} (see Section~\ref{sec:simulations}).

In the case of charm quarks it is a common procedure to adopt Gaussian-smeared interpolating fields~\cite{Gusken:1989qx} for both the source and the sink in order to suppress faster the contribution of the excited states, leading to an improved projection onto the ground state at relatively small time distances. 
For the values of the smearing parameters we set $k_G = 4$ and $N_G = 30$. 
In addition, we apply APE-smearing to the gauge links~\cite{Albanese:1987ds} in the interpolating fields with parameters $\alpha_{APE} = 0.5$ and $N_{APE} = 20$.
The Gaussian smearing is applied as well also for the light and strange quarks. 
The values of $M_\pi$ and $M_K$ reported in Table~\ref{tab:simu&masses} are consistent within the statistical errors with the corresponding ones listed in the Table~1 of Ref.~\cite{Carrasco:2016kpy}, computed using local interpolating fields.

As is well known, at large time distances 2- and 3-point correlation functions behave as
 \bea
        \label{eq:C2_larget}
        C_2^{D(P)}\left(t^\prime,\,\vec{p}_{D(P)}\right) \!\!\!\!& ~ _{\overrightarrow{t^\prime \gg a}} ~ &\!\!\!\! \frac{|Z_{D(P)}|^2}{2E_{D(P)}} 
            \left[ e^{-E_{D(P)} t^\prime} + e^{-E_{D(P)} (T - t^\prime)} \right] , \\
        \label{eq:C3_larget}        
        C^{DP}_{\widehat{\Gamma}}\left(  t,\, t^\prime,\, \vec{p}_D,\, \vec{p}_P \right) \!\!\!\!& ~ _{\overrightarrow{t\gg a\,, \, (t^\prime-t)\gg a}} ~ &\!\!\!\!  
            \frac{Z_P Z_D^*}{4E_P E_D}\, \braket{P(p_P)|\widehat{\Gamma}|D(p_D)}\, e^{-E_D t}\, e^{-E_P (t^\prime - t)}~,
 \eea
where $Z_D$ and $Z_P$ are the matrix elements $\braket{0\lvert\,P_5^D(0)\,\rvert\,D(\vec{p}_D) }$ and $\braket{0\lvert\,P_5^P(0)\,\rvert\,P(\vec{p}_{P})}$, which depend on the meson momenta $\vec{p}_D$ and $\vec{p}_P$ because of the use of smeared interpolating fields, while $E_{D}$ and $E_{P}$ are the energies of the $D$ and $P$ mesons. These energies and matrix elements can be extracted directly by fitting the exponential behavior,  given by the r.h.s of Eq.~(\ref{eq:C2_larget}), of the corresponding 2-point correlation functions. The time intervals $[t_{\rm min},\,t_{\rm max}]$ adopted for the fit (\ref{eq:C2_larget}) are listed in Table~\ref{tab:time_intervals}.
We have checked that the extracted values of $E_{D(P)}$ are nicely reproduced (within the statistical errors) by the continuum-like dispersive relation $E_{D(P)}^{\rm \,disp} = \sqrt{M_{D(P)}^2 + |\vec{p}_{D(P)}|^2}$, where $M_{D(P)}$ is the meson mass extracted from the 2-point correlator corresponding to the meson at rest. 
We decided to use for the analysis the energy values $E_{D(P)}^{\rm \,disp}$ instead of those directly extracted from the fit.

As for the 3-point correlators (\ref{eq:C3_larget}) the usual choice of the time distance $t^\prime$ between the source and the sink is to maximize it, i.e.~to put $t^\prime = T / 2$.
Since we are using smeared interpolating fields, it is convenient to choose smaller values of $t^\prime$, which allow to decrease significantly the statistical noise satisfying at the same time the dominance of the ground-state signals. 
We have optimized the choice of the values of $t^\prime$ for the various lattice spacings and volumes, which can be read off in the last column of Table~\ref{tab:time_intervals}.

\begin{table}[!htb]
\renewcommand{\arraystretch}{1.2} 
\begin{center}
\begin{tabular}{||c|c|c|c||c||}
\hline
$\beta$ & $V / a^4$ &$[t_{\rm min},\,t_{\rm max}]_{(\ell\ell,\,\ell s)}/a$&$[t_{\rm min},\,t_{\rm max}]_{(\ell c)}/a$&$t^\prime/a$\\
\hline
$1.90$ & $32^{3}\times 64$ &$[12,\,31]$&$[8,\,16]$&$18$ \\
& $24^{3}\times 48 $ & $[12,\,23]$&$[8,\,17]$&$18$\\
\hline
$1.95$ & $32^{3}\times 64$ & $[13,\,31]$&$[9,\,18]$&$20$\\
& $24^{3}\times 48 $ &$[13,\,23]$&$[9,\,18]$&$20$\\
\hline
$2.10$ & $48^{3}\times 96$ &$[18,\,40]$&$[12,\,24]$&$26$\\
\hline
\end{tabular}
\end{center}
\renewcommand{\arraystretch}{1.0}
\caption{\it Time intervals adopted for the extraction of the PS meson energies $E_{D(P)}$ and the matrix elements $Z_{D(P)}$ from the 2-point correlators in the light ($\ell$), strange ($s$) and charm ($c$) sectors. The last column contains the values of the time distance $t^\prime$ between the source and the sink adopted for the 3-point correlators (\ref{eq:C3_larget}).}
 \label{tab:time_intervals}
\end{table} 

In our lattice setup we employ maximally twisted fermions and therefore the vector and scalar currents, $\widehat{V}_\mu$ and $\widehat{S}$, renormalize multiplicatively \cite{Frezzotti:2003ni}.
Introducing the local bare currents $V_\mu = \bar{c} \gamma_\mu q$ and $S = \bar{c} q$, where $q = d (s)$, and keeping the same value for the Wilson parameters of the initial and final quarks, one has
  \bea
       \label{eq:Vmu}
       \widehat{V}_\mu \!\!\!\! & = & \!\!\!\! {\cal{Z}}_V \cdot V_\mu = {\cal{Z}}_V \, \bar{c} \gamma_\mu q ~,\\
       \label{eq:S}
       \widehat{S} \!\!\!\! & = & \!\!\!\! {\cal{Z}}_P \cdot S = {\cal{Z}}_P \, \bar{c} q ~ ,
 \eea
where ${\cal{Z}}_V$ and ${\cal{Z}}_P$ are the renormalization constants (RCs) of the vector and pseudoscalar densities for standard Wilson fermions, respectively.
The twisted quark masses renormalize multiplicatively with a RC ${\cal{Z}}_m$ given by ${\cal{Z}}_m = 1 / {\cal{Z}}_P$ \cite{Frezzotti:2003ni}, which means that the product $(m_c - m_q) \braket{\widehat{S}}$ is renormalization group invariant.
Therefore, according to Eq.~(\ref{eq:scalar_form_factor_2}), the scalar form factor $f_0(q^2)$ is related to the (bare) matrix element $\braket{S}$ by
 \be
      \label{eq:f0_S}
      \braket{S} \equiv \braket{P(p_P) | S | D(p_D)} = \frac{M_D^2 - M_P^2}{\mu_c - \mu_q}\, f_0(q^2) ~ ,
 \ee
where $\mu_q$ and $\mu_c$ are the bare light (strange) and charm quark masses, respectively. 

As in the case of the semileptonic $K_{\ell 3}$ decay~\cite{Carrasco:2016kpy}, the matrix elements $\langle S \rangle$  and $\langle \widehat{V}_\mu \rangle$ (see Eq.~(\ref{eq:matrix_element_decomposition})) can be extracted from the time dependence of the ratios $R_\mu$ $(\mu=0,1,2,3)$ and $R_S$ of 2- and 3-point correlation functions, defined as in Eqs.~(\ref{eq:C2}-\ref{eq:C3}) but using the local bare currents $V_\mu$ and $S$, namely
 \bea
    \label{eq:Rmu} 
    R_\mu(t,\vec{p}_D, \vec{p}_P) \!\!\!&\equiv&\!\!\! 4\, p_{D \mu}\, p_{P \mu}\, \frac{C^{DP}_{V_\mu}(t, t^\prime, \vec{p}_D, \vec{p}_P) \, 
        C^{P D}_{V_\mu}(t, t^\prime, \vec{p}_P, \vec{p}_D)} {C^{P P}_{V_\mu}(t, t^\prime, \vec{p}_P, \vec{p}_P) \, 
        C^{DD}_{V_\mu}(t, t^\prime, \vec{p}_D, \vec{p}_D)} ~ ,\\[2mm]
    \label{eq:RS}
    R_S(t,\vec{p}_D, \vec{p}_P) \!\!\!&\equiv&\!\!\! 4\, E_D\, E_P\,\frac{C^{DP}_S(t, t^\prime,\vec{p}_D, \vec{p}_P)\, C^{P D}_S(t, t^\prime,\vec{p}_P, \vec{p}_D)} {\widetilde{C}_2^D \left(t^\prime,\vec{p}_{D} \right)\, \widetilde{C}_2^P \left(t^\prime,\vec{p}_{P} \right)} ~ ,
 \eea
where the correlation functions $\widetilde{C}_2^{D(P)}(t)$ are given by
 \be
      \widetilde{C}_2^{D(P)} \left(t,\,\vec{p}_{D(P)}\right) \equiv \frac{1}{2}\left[ C_2^{D(P)}\left(t,\,\vec{p}_{D(P)}\right) + 
          \sqrt{C_2^{D(P)}\left(t,\,\vec{p}_{D(P)}\right)^2 - C_2^{D(P)}\left(\frac{T}{2},\,\vec{p}_{D(P)}\right)^2} \right] ~ ,
      \label{eq:C2_tilde}
 \ee
which at large time distances behave as
 \be
      \widetilde{C}_2^{D(P)} \left(t,\,\vec{p}_{D(P)}\right) ~ _{\overrightarrow{t \gg a}} ~ \frac{Z_{D(P)}}{2\, E_{D(P)}}\,e^{-E_{D(P)} t} ~ ,
     \label{eq:C2_tilde_larget}
 \ee
i.e.~without the backward signal.
Note that the denominator of Eq.~(\ref{eq:Rmu}) is nothing but the numerator evaluated in the mass-degenerate limit for the valence quarks in the current. 
Such mass-degenerate quarks have the same lattice regularisation of the corresponding ones in the numerator, so that the RC ${\cal{Z}}_V$ is the same for the two terms in the ratio and therefore it cancels out.

At large time distances one has:
\bea
\label{eq:Rmu_plateau}
R_\mu(t,\vec{p}_D, \vec{p}_P) \!\!\!\!&~ _{\overrightarrow{t\gg a\, \, (t^\prime-t)\gg a}} ~&\!\!\!\! 4\, p_{D \mu}\, p_{P \mu}\,\frac{\braket{P(p_P) | \widehat{V}_\mu |D(p_D)}\braket{D(p_D) | \widehat{V}_\mu |P(p_P)}}{\braket{P(p_P) | \widehat{V}_\mu |P(p_P)}\braket{D(p_D) | \widehat{V}_\mu |D(p_D)}} =\lvert \braket{\widehat{V}_\mu} \rvert^2~ ,\\
\nonumber \\
\label{eq:RS_plateau}
R_S(t,\vec{p}_D, \vec{p}_P) \!\!\!\!&~ _{\overrightarrow{t\gg a\, \, (t^\prime-t)\gg a}} ~&\!\!\!\! \lvert\braket{P(p_P)| S |D(p_D)}\rvert^2 = \lvert\braket{S}\rvert^2~,
\eea
so that, up to lattice artifacts, the renormalized quantity $|\braket{\widehat{V}_\mu}|^2$ and the bare one $|\braket{S}|^2$ can be extracted directly from the plateau of $R_\mu$ and $R_S$, independently of the matrix  elements $Z_D$ and $Z_{\pi(K)}$ of the interpolating fields.
In the r.h.s.~of Eq.~(\ref{eq:Rmu_plateau}) we have used the fact that, due to the charge conservation, $\braket{P(p_P)|\widehat{V}_\mu|P(p_P)}=2\,p_{P\mu}$ and $\braket{D(p_D)|\widehat{V}_\mu|D(p_D)}=2\,p_{D\mu}$.
Taking the  square root of $R_\mu$ and $R_S$ we can get the absolute value of the matrix elements $\braket{\widehat{V}_\mu}$ and $\braket{S}$, while their sign can be easily inferred from those of the correlators $C_{V_\mu}^{DP}(t, t^\prime, \vec{p}_D, \vec{p}_P)$ and $C_S^{DP}(t, t^\prime, \vec{p}_D, \vec{p}_P)$ in the relevant time regions.

When a spatial component of the momentum of either the parent or the child meson is vanishing, the corresponding matrix element $\braket{P(p_P) | V_i | P(p_P)}$ (or $\braket{D(p_D) | V_i | D(p_D)}$) ($i = 1, 2, 3$) is also vanishing and the correlator $C^{P P}_{V_i}$ (or $C^{DD}_{V_i}$), cannot be used in the denominator of the r.h.s~of Eq.~(\ref{eq:Rmu}).
In these cases the quantity $2p_{P i}\,/\! \braket{P(p_P) | V_i | P(p_P)}$ (or $2p_{D i}\,/\!\braket{D(p_D) | V_i | D(p_D)}$) is replaced by $2E_P\,/ C^{P P}_{V_0}$ (or $2E_D\,/ C^{DD}_{V_0}$).

When both the pion (kaon) and the D-meson are at rest, only two ratios, $R_S$ and $R_0$, can be constructed, providing two independent determinations of the scalar form factor $f_0(q^2)$ at the kinematical end-point $q^2_{\rm max}=\left( M_D-M_P \right)^2$, which may differ by lattice artifacts.

In order to inject momenta on the lattice we use the same procedure adopted in Ref.~\cite{Carrasco:2016kpy} for the $K_{\ell 3}$ decays. 
In particular for the valence quark fields we impose twisted boundary conditions (BC's) \cite{Bedaque:2004kc,deDivitiis:2004kq,Guadagnoli:2005be} in the spatial directions and anti-periodic BC's in time. 
The sea dynamical quarks, on the contrary, have been simulated with periodic BC's in space and anti-periodic ones in time. 
The choice of using twisted BC's for the valence quark fields is crucial in order to remove the strong limitations, resulting from the use of periodic BC's, to the accessible kinematical regions of momentum-dependent quantities like, e.g., form factors.
Furthermore we remark that, as shown in Refs.~\cite{Sachrajda:2004mi,Bedaque:2004ax}, for physical quantities which do not involve final state interactions (like, e.g., meson masses, decay constants and semileptonic form factors), the use of different BC's for valence and sea quarks produces only finite size effects (FSEs) which are exponentially small.

The quark 3-momentum is then given by
 \be
    p = \frac{2 \pi}{L}\left( \theta + n_x , ~ \theta + n_y ,~ \theta + n_z \right) ~ ,
    \label{eq:pj_twisted}
  \ee
where $n_{x,y,z}$ are integers and the $\theta$ values, adopted for the different gauge ensembles and democratically distributed along the three spatial directions, are collected in Table~\ref{tab:theta_values}. 
They have been chosen in order to obtain momenta with values ranging from $\approx 150\MeV$ to $\approx 650\MeV$ for all the various lattice spacings and volumes\footnote{The correlators used in this work have been calculated within the PRACE project PRA067 {\it ``First Lattice QCD study of B-physics with four flavors of dynamical quarks"}. 
The values of the quark momentum were not chosen having in mind the investigation of hypercubic effects in the semileptonic form factors. In particular the use of spatially symmetric values of the quark momentum (see Table \ref{tab:theta_values}) is not ideal for such a purpose.}.
\begin{table}[!htb] 
\renewcommand{\arraystretch}{1.2} 
\begin{center}
\begin{tabular}{||c|c||c||}
\hline
$\beta$ & $V / a^4$ & $\theta$\\
\hline
$1.90$ & $32^{3} \times 64$ & $0.0, ~ \pm 0.200, ~ \pm0 .467, ~ \pm 0.867$\\
\cline{2-3}
            & $24^{3} \times 48 $ & $0.0, ~ \pm 0.150, ~ \pm 0.350, ~ \pm 0.650$\\
\hline
$1.95$ & $32^{3} \times 64$ & $0.0, ~ \pm 0.183, ~ \pm 0.427, ~ \pm 0.794$\\
\cline{2-3}
            & $24^{3} \times 48 $ &$0.0, ~ \pm 0.138, ~ \pm 0.321, ~ \pm 0.596$\\
\hline
$2.10$ & $48^{3} \times 96$ &$0.0, ~ \pm 0.212, ~ \pm 0.493, ~ \pm 0.916$\\
\hline
\end{tabular} 
\end{center}
\renewcommand{\arraystretch}{1.0} 
\caption{\it Values of the parameter $\theta$, appearing in Eq.~(\ref{eq:pj_twisted}), for the various ETMC gauge ensembles of Table~\ref{tab:simu&masses}.}
\label{tab:theta_values}
\end{table} 

The 3-point correlation functions $C^{DP}_{V_\mu}(t, t^\prime, \vec{p}_D, \vec{p}_P)$ and $C^{DP}_S(t, t^\prime, \vec{p}_D, \vec{p}_P)$ have been simulated imposing periodic BC's to the spectator light quark and partially twisted BC's (\ref{eq:pj_twisted}) to the initial $c$ and final $u(s)$ quarks. 
With this choice the $\pi$, $K$ and $D$ meson (spatial) momenta are given by $\vec{p}_D = (2 \pi / L) \left( \theta_D, \theta_D, \theta_D \right)$ and $\vec{p}_{\pi(K)} = (2 \pi / L) \left( \theta_{\pi(K)}, \theta_{\pi(K)}, \theta_{\pi(K)} \right)$, where $\theta_D$ and $\theta_{\pi(K)}$ can assume for each gauge ensemble the values of the parameter $\theta$ given in Table~\ref{tab:theta_values}. 

As described in Ref.~\cite{Frezzotti:2003ni} the use of two kinematics with opposite spatial momenta of the initial and final mesons, given by opposite signs of the corresponding $\theta$, allows for an ${\cal{O}}(a)$ improvement on the matrix elements $\braket{\widehat{V}_\mu}$ and $\braket{S}$ performing the following average:
 \bea
       \label{eq:improvement_V0}
       \braket{\widehat{V}_0}_{\rm imp} & \equiv & \frac{1}{2} \left[ \braket{P(E_P, \vec{p}_P) | \widehat{V}_0 | D(E_D, \vec{p}_D)} + 
           \braket{P(E_P, -\vec{p}_P) | \widehat{V}_0 | D(E_D, -\vec{p}_D)} \right] ~ , \\
       \label{eq:improvement_Vi}
       \braket{\widehat{V}_i}_{\rm imp} & \equiv & \frac{1}{2} \left[ \braket{P(E_P, \vec{p}_P) | \widehat{V}_i | D(E_D, \vec{p}_D)} - 
           \braket{P(E_P, -\vec{p}_P) | \widehat{V}_i | D(E_D, -\vec{p}_D)} \right] ~ , \\
       \label{eq:improvement_VS}
       \braket{S}_{\rm imp} & \equiv & \frac{1}{2} \left[ \braket{P(E_P, \vec{p}_P) | S | D(E_D, \vec{p}_D)} +
            \braket{P(E_P, -\vec{p}_P) | S | D(E_D, -\vec{p}_D)} \right] ~ . 
\eea
Furthermore since we are using democratically distributed momenta in the three spatial directions, the matrix elements of the spatial components of the vector current $\braket{\widehat{V}_i}_{\rm imp}$ are equal to each other. 
Therefore, in order to improve the statistics, we average them to get
 \be
       \label{eq:Vsp}
       \braket{\widehat{V}_{\rm sp}}_{\rm imp} \equiv \frac{1}{3} \left[ \braket{\widehat{V}_1}_{\rm imp} + \braket{\widehat{V}_2}_{\rm imp} + 
                                                                                 \braket{\widehat{V}_3}_{\rm imp} \right]~.
 \ee

\begin{figure}[htb!]
\centering{
\includegraphics[scale=0.60]{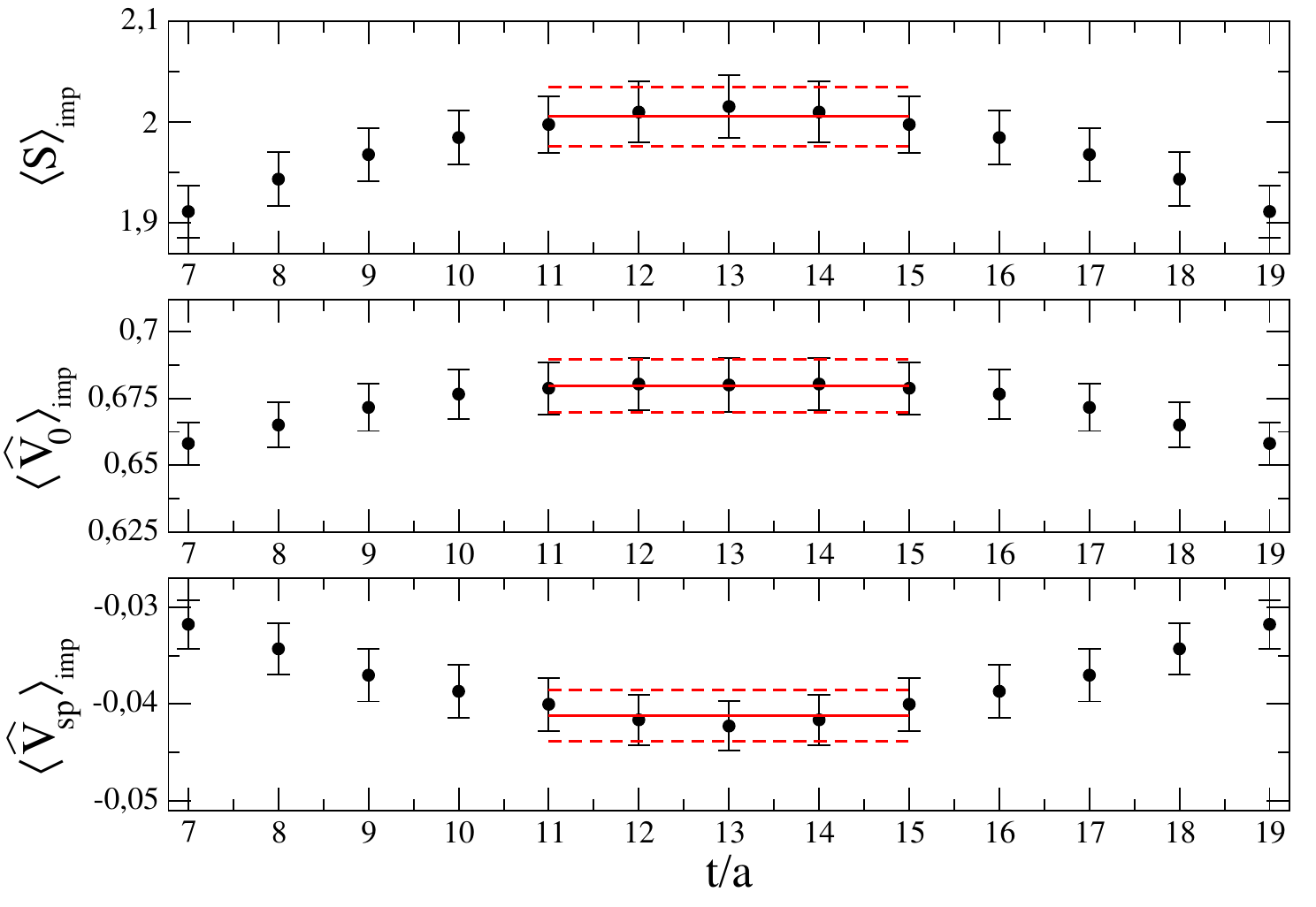}
}
\caption{\it Matrix elements $\braket{\widehat{V}_{\rm sp}}_{\rm imp}$, $\braket{\widehat{V}_0}_{\rm imp}$ and $\braket{S}_{\rm imp}$ for the $D\to\pi$ case extracted from the ratios $R_{\mu}$ and $R_S$ [see Eqs.~(\ref{eq:Rmu_plateau}) and (\ref{eq:RS_plateau})] for the ensemble D20.48 with $\beta = 2.10$, $L / a = 48$, $a\mu_\ell = 0.0020$, $a\mu_c = 0.17$, $\vec{p}_D = - \vec{p}_\pi$ and $|\vec{p}_D| \simeq 150\MeV$. The meson masses are $M_\pi \simeq 254 \MeV$  and $M_D \simeq 1933\MeV $. The horizontal red lines correspond to the plateau regions used to extract the matrix elements and to their central values and statistical errors.}
\label{fig:plateau}
\end{figure}

The quality of the plateau for the matrix elements $\braket{\widehat{V}_0}_{\rm imp}$, $\braket{\widehat{V}_{\rm sp}}_{\rm imp}$ and $\braket{S}_{\rm imp}$ is illustrated in Fig.~\ref{fig:plateau} in the case of the $D \to \pi$ transition.
The time intervals adopted for fitting Eqs.~(\ref{eq:Rmu_plateau}-\ref{eq:RS_plateau}) are symmetric around $t^\prime / 2$ (see Table~\ref{tab:time_intervals} for the values of $t^\prime$ for each specific gauge ensemble) and equal to $[t^\prime / 2 - 2, ~ t^\prime / 2 + 2]$.
These values are compatible with the dominance of the $\pi$, $K$ and $D$ mesons ground-state observed along the time intervals of Table~\ref{tab:time_intervals} for the two-point correlation functions.

Thus, from the 2- and 3-point lattice correlators we are able to extract the three ${\cal{O}}(a)$-improved matrix elements $\braket{\widehat{V}_0}_{\rm imp}$, $\braket{\widehat{V}_{\rm sp}}_{\rm imp}$ and $\braket{S}_{\rm imp}$. 
The standard procedure for determining the scalar and vector form factors $f_0(q^2)$ and $f_+(q^2)$ is to assume the following Lorentz-covariant decomposition
 \bea
     \label{eq:vectorL}
     \braket{P(p_P) | \widehat{V}_\mu | D(p_D)} & = & P_\mu ~ f_+(q^2) + q_\mu ~ \frac{M_D^2 - M_P^2}{q^2} \left[ f_0(q^2) - f_+(q^2) \right] +
                                                                                    {\cal{O}}(a^2) ~ , \\[2mm]
     \label{eq:scalarL}
     \braket{P(p_P) | S | D(p_D)} & = & \frac{M_D^2 - M_P^2}{\mu_c - \mu_q} f_0(q^2) + {\cal{O}}(a^2)
  \eea
with $P_\mu \equiv (p_D + p_P)_\mu$.
Explicitly one has
 \bea
       \label{eq:V0_final}
       \braket{\widehat{V}_0}_{\rm imp} & = & (E_D + E_P) f_+(q^2) + (E_D - E_P) \frac{M_D^2 - M_P^2}{q^2} \left[ f_0(q^2) - f_+(q^2) \right] + 
                                                                      {\cal{O}}(a^2) ~ , \\[2mm]
       \label{eq:Vsp_final}
       \braket{\widehat{V}_{\rm sp}}_{\rm imp} & = & \frac{2 \pi}{L} \left\{ (\theta_D + \theta_P) f_+(q^2) + (\theta_D - \theta_P) \frac{M_D^2 - M_P^2}{q^2} 
                                                                                \left[ f_0(q^2) - f_+(q^2) \right] \right\} + {\cal{O}}(a^2) ~ , \qquad \\[2mm]
       \label{eq:S_final}
       \braket{S}_{\rm imp} & = & \frac{M_D^2 - M_P^2}{\mu_c - \mu_q} f_0(q^2) + {\cal{O}}(a^2) ~ ,
 \eea
which represent a redundant mathematical system consisting of three equations depending on just two form factors.

We then determine $f_0(q^2)$ and $f_+(q^2)$ by minimizing the $\chi^2$-variable constructed using Eqs.~(\ref{eq:V0_final}-\ref{eq:S_final}).
In the next Section we present and discuss the result of this determination in which, as anticipated, we found evidence for Lorentz symmetry breaking terms.

\section{Form factors and hypercubic effects}
\label{sec:sec2}

\begin{figure}[htb!]
\centering{
\includegraphics[scale=0.30]{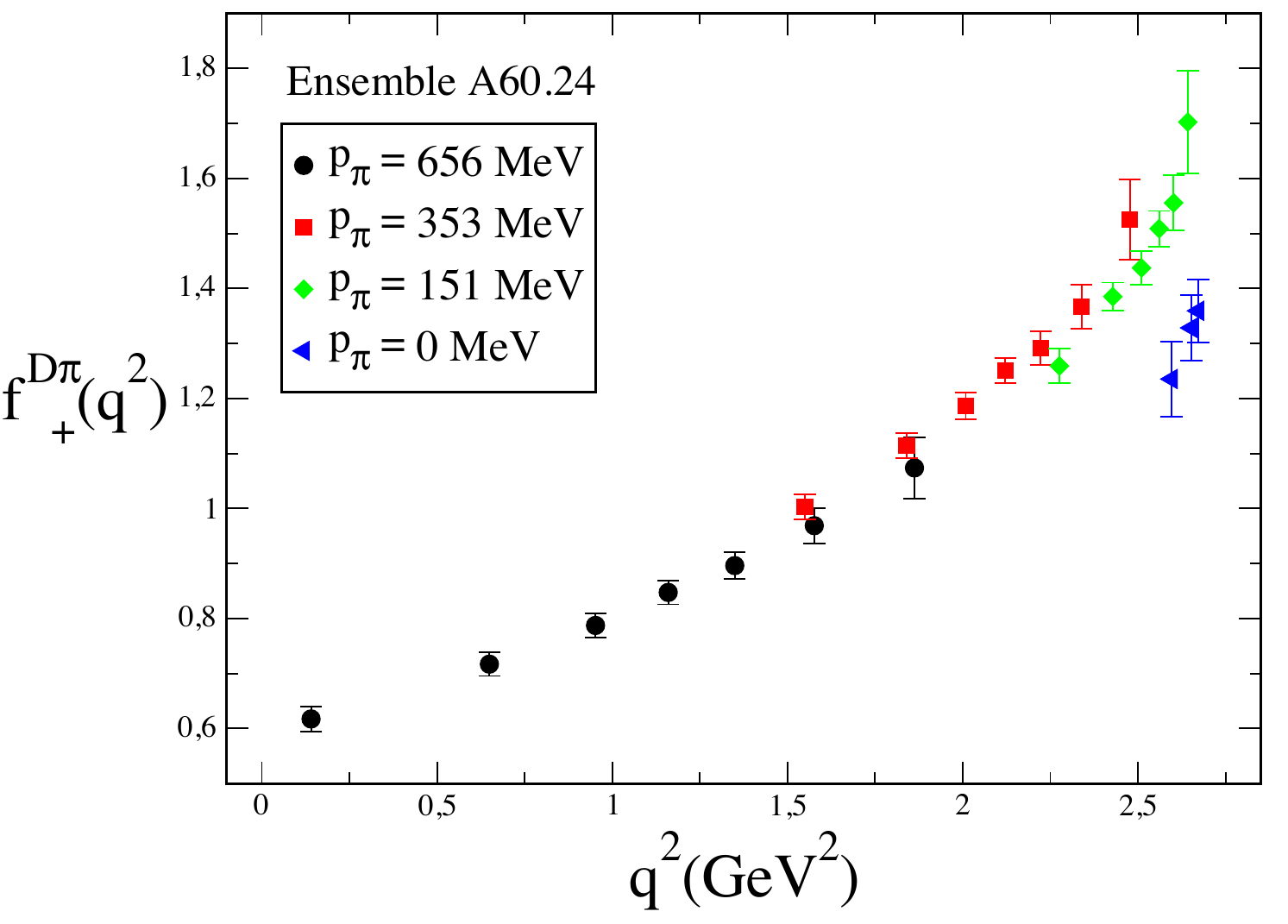} ~~ 
\includegraphics[scale=0.30]{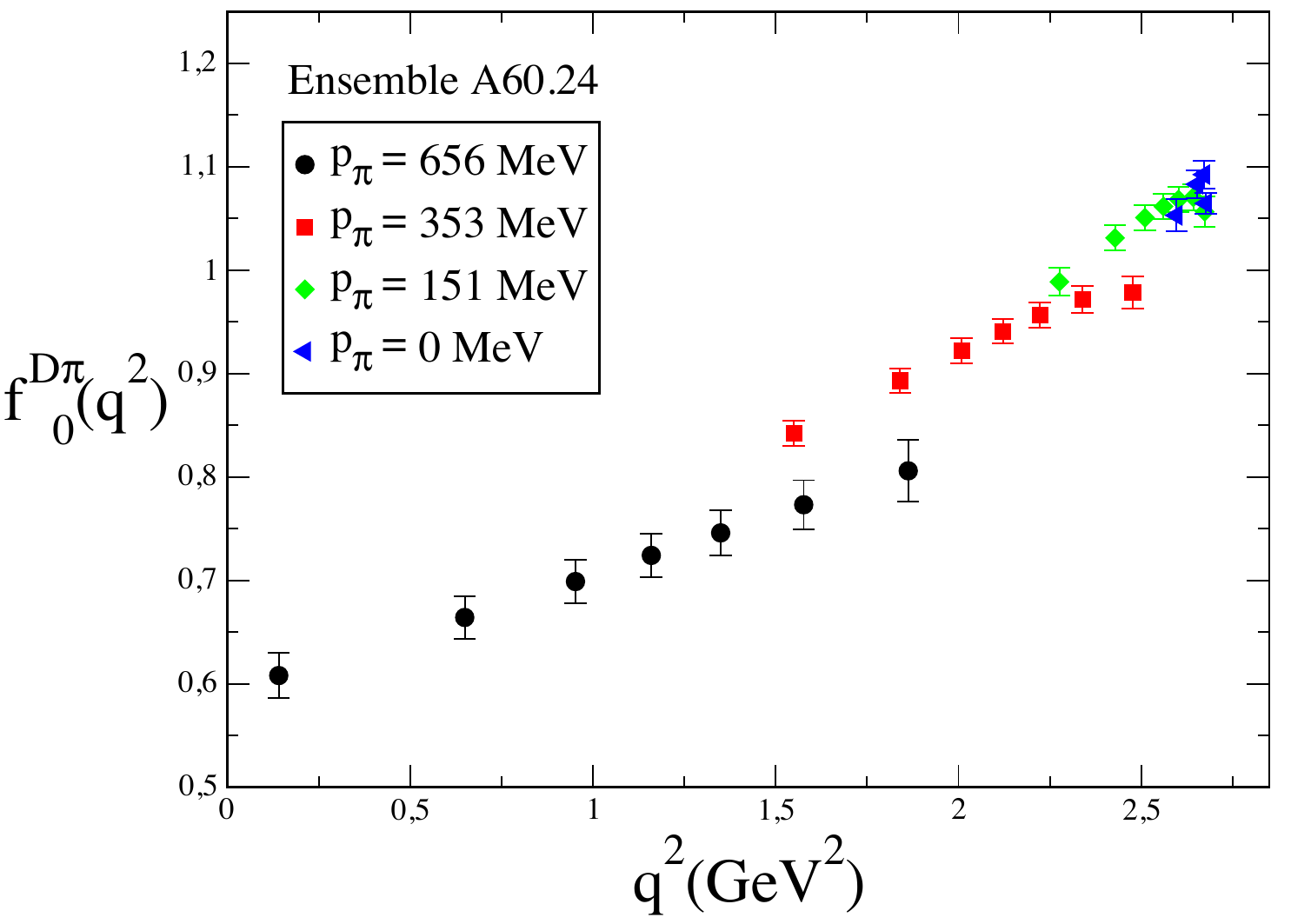}
\includegraphics[scale=0.30]{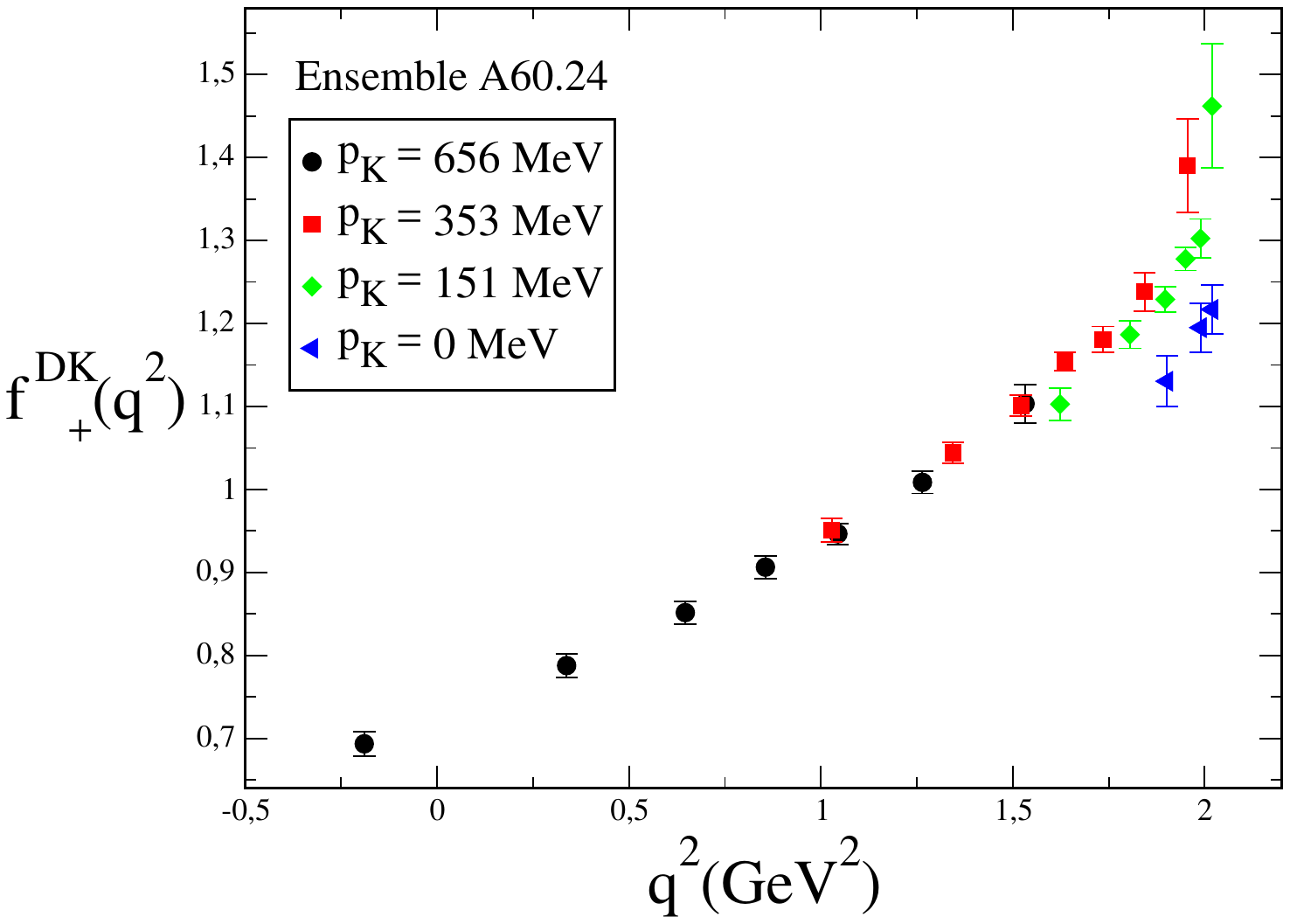} ~~
\includegraphics[scale=0.30]{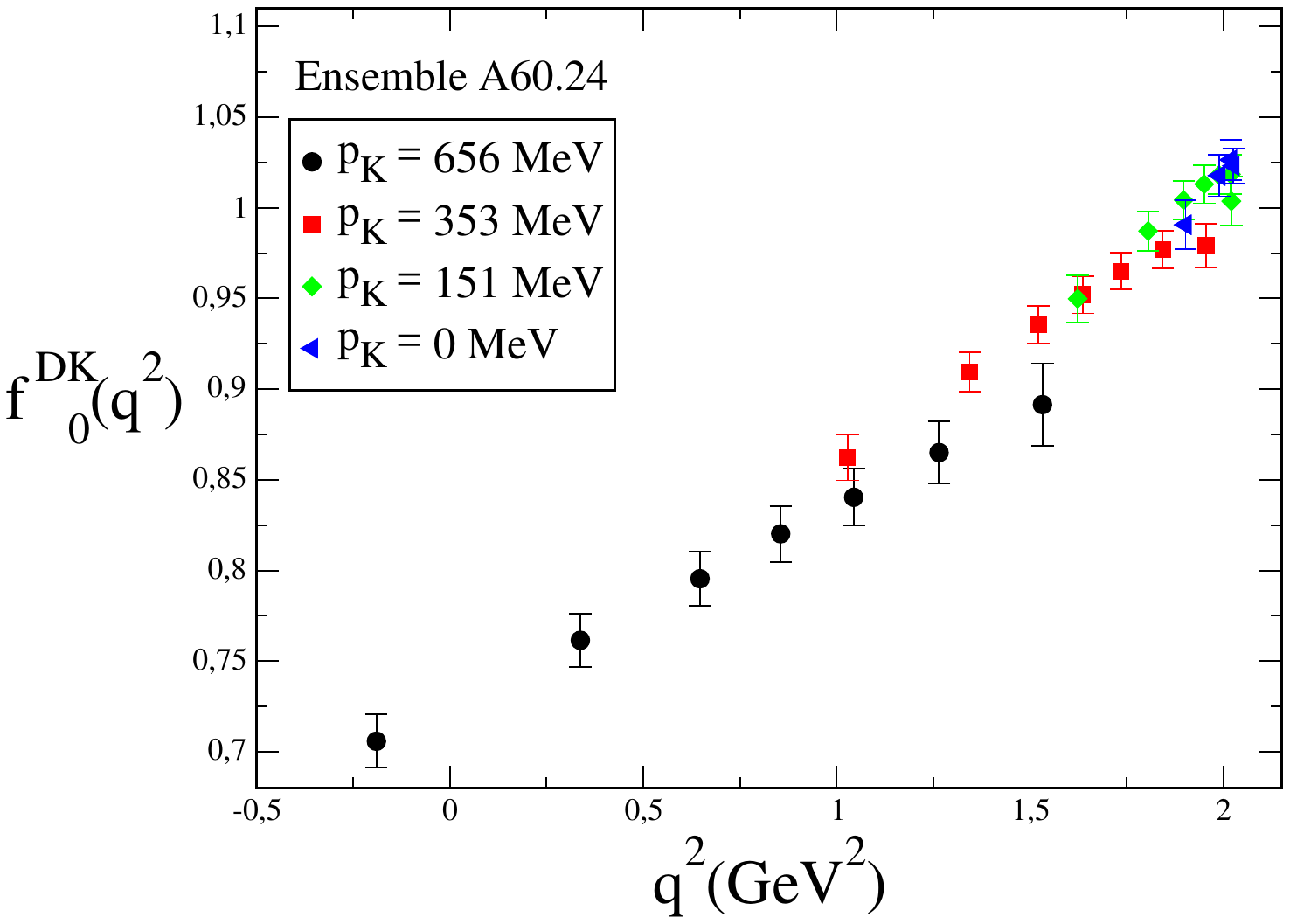}
}
\caption{\it Momentum dependence of the vector $f_+^{D\pi}$ (upper left panel), $f_+^{D K}$ (lower left panel) and scalar $f_0^{D\pi}$ (upper right panel), $f_0^{D K}$ (lower right panel) form factors in the case of the gauge ensemble A60.24 \protect\cite{Carrasco:2014cwa}. Different markers and colors distinguish different values of the child meson momentum. The simulated meson masses are $M_\pi \simeq 386$ MeV, $M_K \simeq 599$ MeV and $M_D \simeq 2022$ MeV, while the lattice spacing and spatial size are $a \simeq 0.0885$ fm and $L \simeq 2.13$ fm, respectively (see Table \ref{tab:simu&masses}).}
\label{fishbone}
\end{figure}

After a small interpolation of our lattice data to the physical values of the strange and charm quark masses, $m_s^{phys}(2\,\rm{GeV})=99.6\,(4.3)\,$MeV and $m_c^{phys}(2\,\rm{GeV})=1.176\,(3.9)\,$GeV taken from Ref.~\cite{Carrasco:2014cwa}, we determine the vector and scalar form factors $f_+^{D \pi(K)}$ and $f_0^{D \pi(K)}$ for each gauge ensemble and for each choice of parent and child meson momenta. 
The momentum dependencies of the semileptonic form factors $f_+^{D \pi(K)}$ and $f_0^{D \pi(K)}$ are illustrated in the upper (lower) panels of Fig.~\ref{fishbone}, where different markers and colors correspond to different values of the child meson momentum for the ETMC ensemble A60.24, i.e.~at fixed values of the parent and child meson masses as well as of the lattice spacing and volume.
Therefore, if the Lorentz-covariant decomposition (\ref{eq:vectorL}-\ref{eq:scalarL}) were adequate to describe our data, the extracted form factors would depend only on the squared 4-momentum transfer $q^2$.
This is not the case and an extra dependence on the value of the child meson momentum is clearly visible in Fig.~\ref{fishbone} beyond the statistical uncertainties.  

As is well known, the lattice breaks Lorentz symmetry and is invariant only under discrete rotations by multiple of $90^\circ$ in each direction of the Euclidean space-time.
Therefore, the matrix elements (\ref{eq:improvement_V0}-\ref{eq:Vsp}), and consequently the form factors extracted from Eqs.~(\ref{eq:V0_final}-\ref{eq:S_final}), may depend also on hypercubic invariants\footnote{Hypercubic symmetry is also broken on our lattices because of the different length of the temporal and spatial dimensions. This effect however is expected to be subdominant and will be neglected in what follows.}.
Hypercubic effects are known to affect lattice calculations and they have been discussed for instance in Refs.~\cite{Becirevic:1999uc,deSoto:2007ht}.
It is however the first time that these effects are observed in the $D$-meson semileptonic form factors. 
In Refs.~\cite{Aubin:2004ej} and \cite{Na:2011mc,Na:2010uf} $N_f = 2 + 1$ results for the $f_+^{D \pi(K)}$ form factor have been obtained by FNAL/MILC and HPQCD collaborations, respectively, using only a limited number of kinematic conditions, restricted in particular to the parent $D$-meson at rest.
Also the ETMC reported $N_f = 2$ results for the $D$-meson semileptonic form factors in Ref.~\cite{DiVita:2011py}, but the kinematics were limited to the Breit-frame ($\vec{p}_D = - \vec {p}_P$). 
Recently in Ref.~\cite{Kaneko:2017sct} both the $D \to \pi$ and $D \to K$ semileptonic transitions have been investigated using $N_f = 2+1$ domain-wall fermions assuming the $D$-meson at rest, while in Ref.~\cite{Primer:2017xzs} the FNAL/MILC collaboration has addressed the determination of the semileptonic form factors, using $N_f = 2+1+1$ MILC ensembles with HISQ fermions and tuning properly the child meson momentum to reach directly $q^2 = 0$, but working only in the reference frame where the $D$-meson is at rest.
We argue that all these choices may obscure the presence of hypercubic effects in the lattice data. 

\begin{figure}[htb!]
\centering{
\includegraphics[scale=0.30]{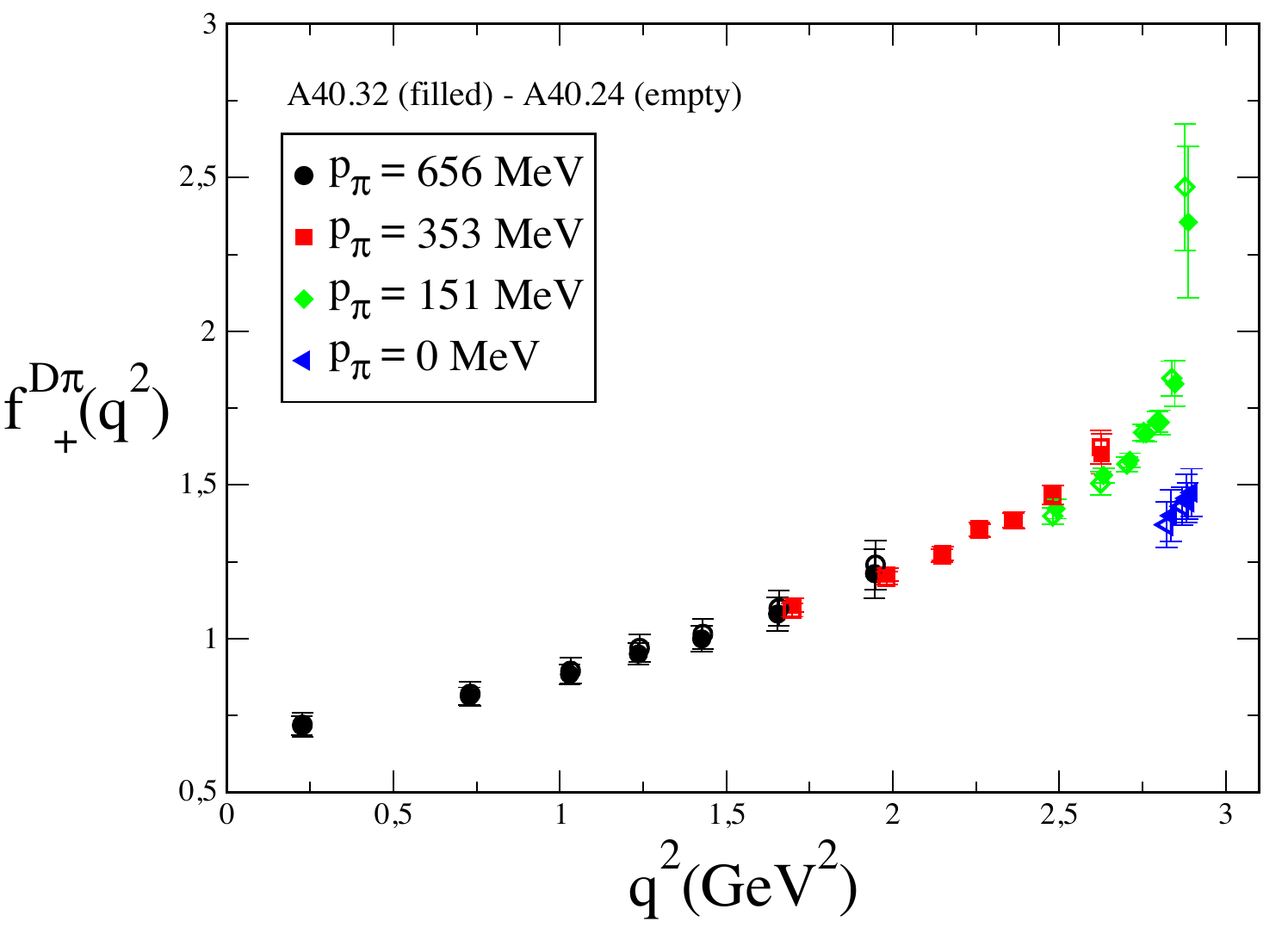} ~~ 
\includegraphics[scale=0.30]{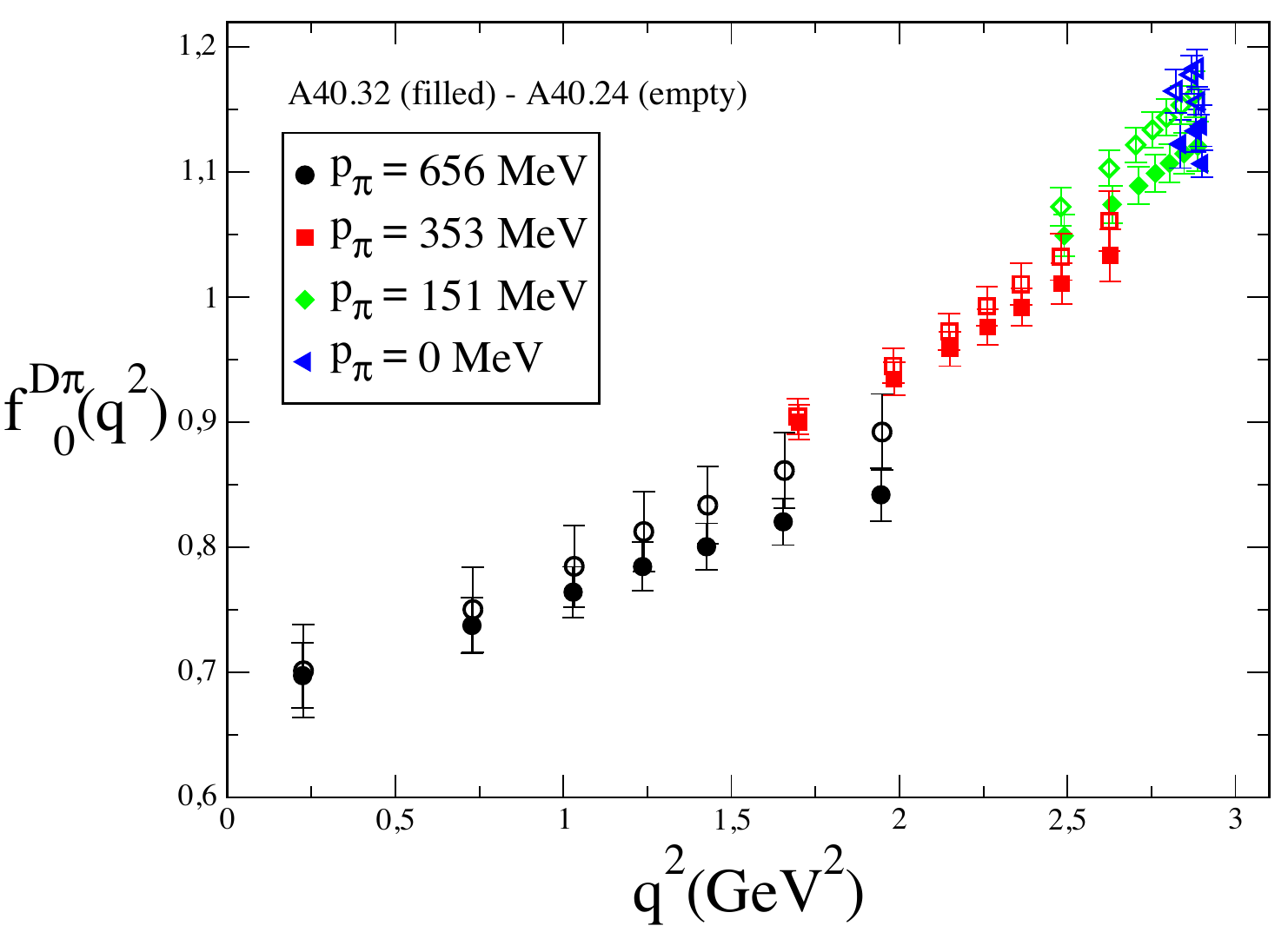}
\includegraphics[scale=0.30]{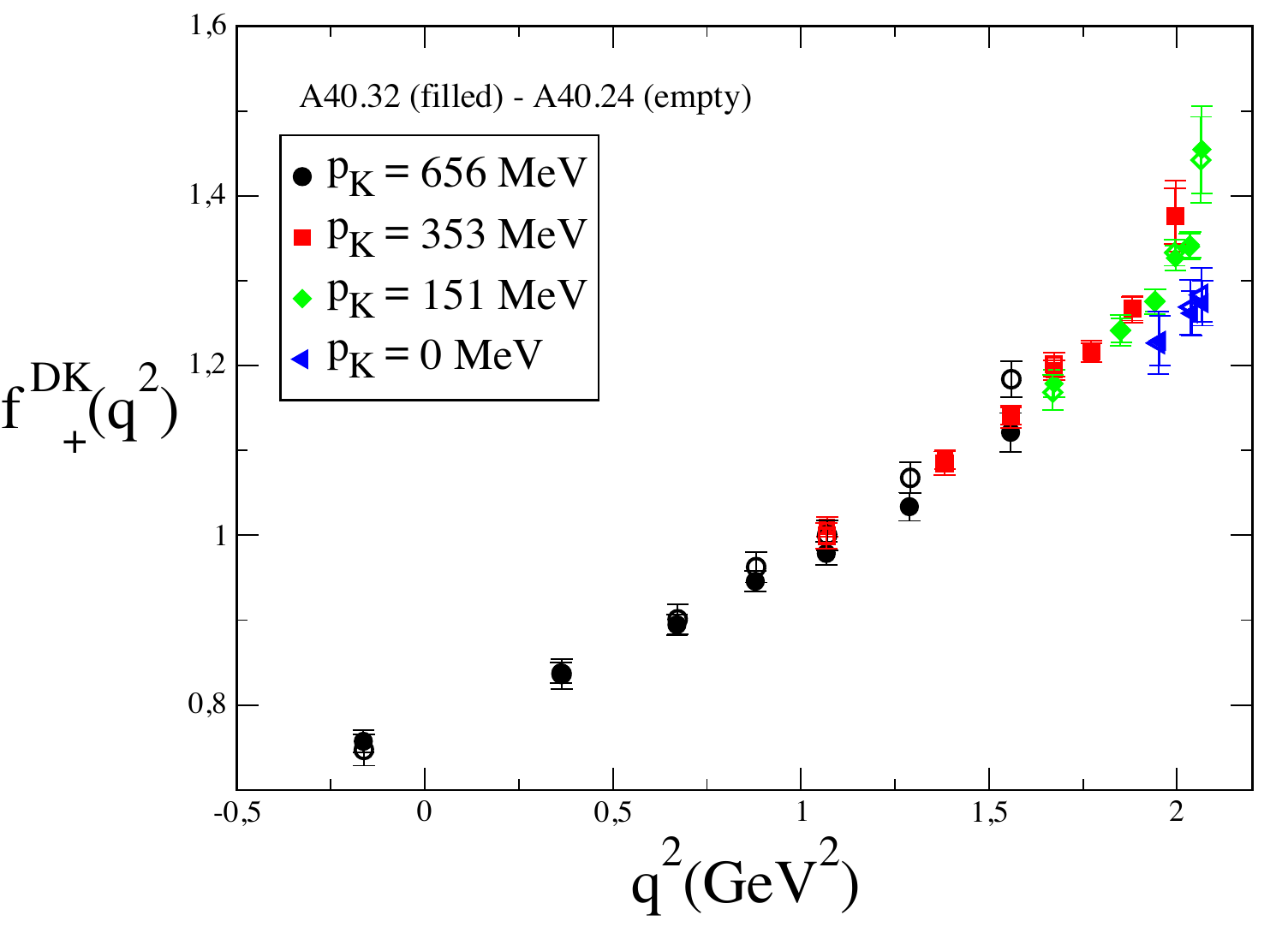} ~~~~ 
\includegraphics[scale=0.30]{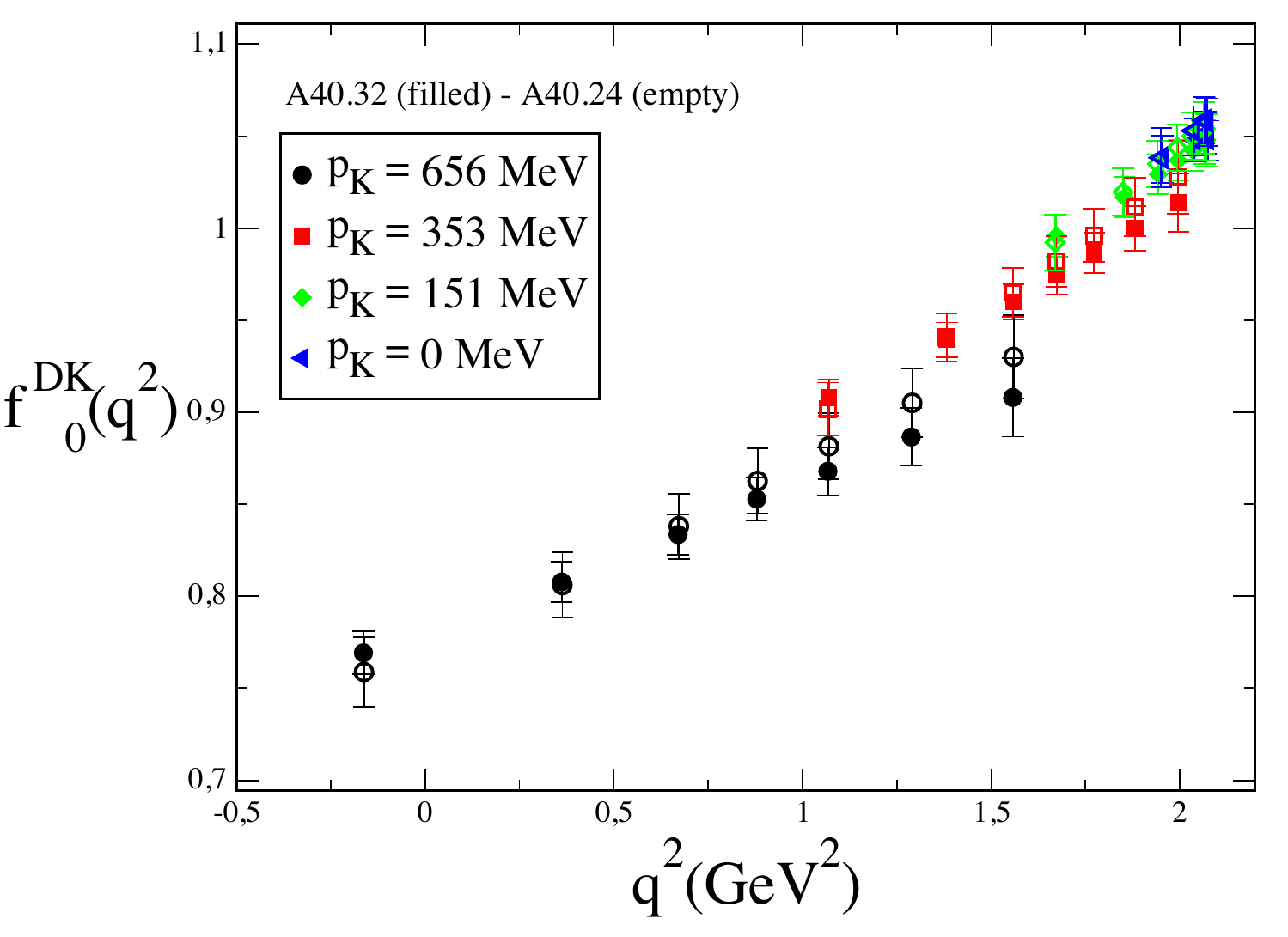}
}
\caption{\it Results for the vector (left panels) and scalar (right panels) form factors in the case of the $D \to \pi$ (upper panels) and $D \to K$ (lower panels) semileptonic decays versus $q^2$ for the gauge ensembles A40.24 and A40.32, which correspond to $a \simeq 0.0885$ fm, $M_\pi \simeq 320$ MeV, $M_K \simeq 580$ MeV, $M_D \simeq 2020$ MeV and two different lattice volumes $L / a = 24$ (empty markers) and $L / a = 32$ (filled markers), respectively. The different shape and color of the markers distinguish between different values of the child meson momentum.}
\label{fig:FSE}
\end{figure}

The behavior observed in Fig.~\ref{fishbone} might be (at least partially) related to finite size effects (FSEs) (see, e.g., Ref.~\cite{Bernard:2017scg} for the case of $K_{\ell 3}$ decays). 
The possible impact of FSEs has been investigated by comparing the results corresponding to the two gauge ensembles, A40.24 and A40.32, which share the same pion mass and lattice spacing, but have different lattice sizes, $L = 24 a$ and $L = 32 a$, respectively, as illustrated in Fig.~\ref{fig:FSE}. 
It can clearly be seen that for the $D \to \pi(K)$ semileptonic vector and scalar form factors FSEs appear to be negligible within the current statistical uncertainties, except for the slope of the $D \to \pi$ scalar form factor (upper right panel in Fig.~\ref{fig:FSE}).
Hypercubic effects for the two gauge ensembles A40.24 and A40.32 are found to be comparable and they do not appear to depend on the lattice size $L$.

Thus, since in our setup all the current matrix elements are ${\cal{O}}(a)$-improved, the breaking of the Lorentz invariance is expected to be produced by ${\cal{O}}(a^2)$ hypercubic effects, whose subtraction will be discussed in the next Section in order to get the Lorentz-invariant semileptonic vector and scalar form factors.

\begin{figure}[htb!]
\centering{
\includegraphics[scale=0.30]{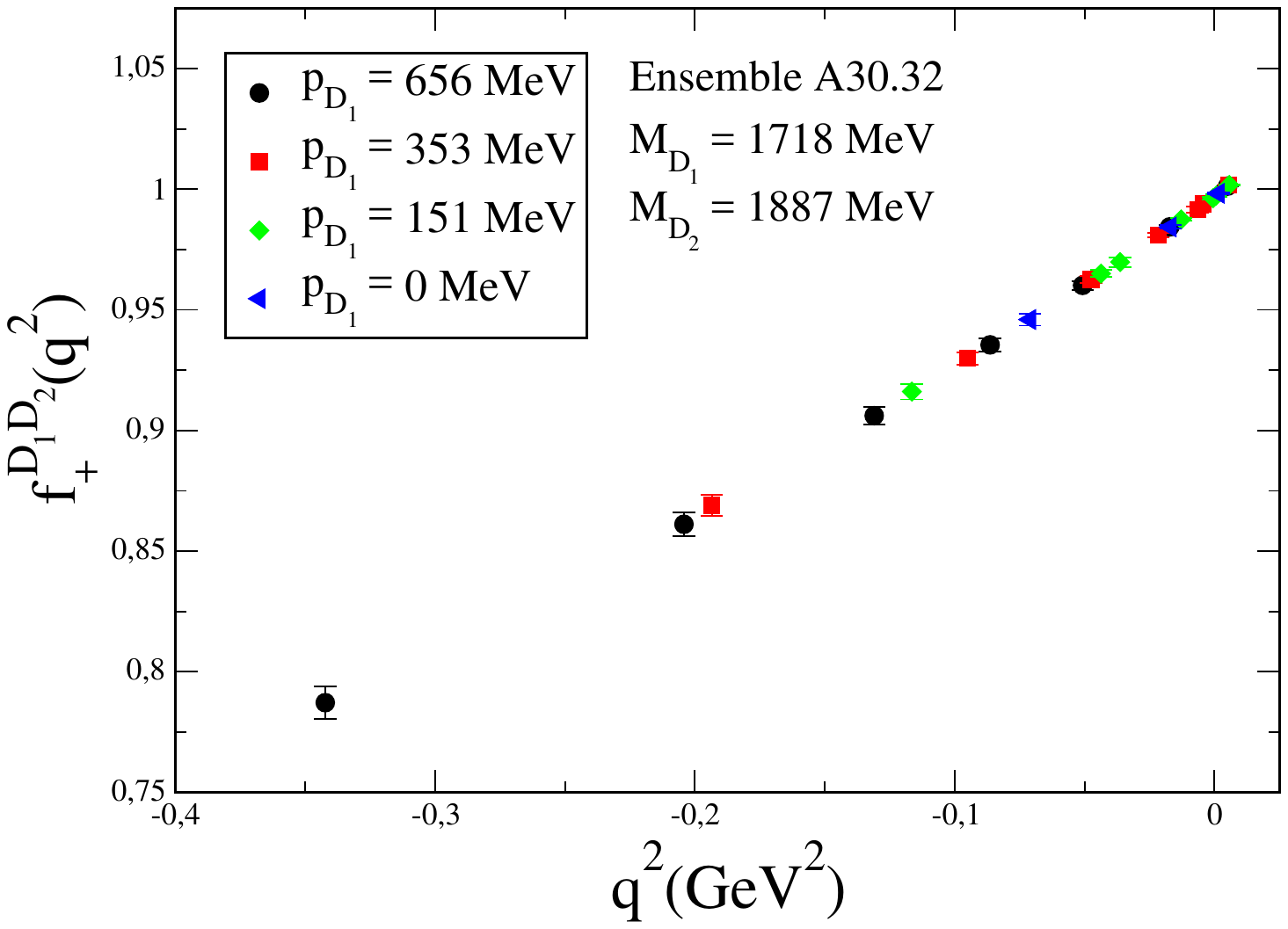} ~~ 
\includegraphics[scale=0.30]{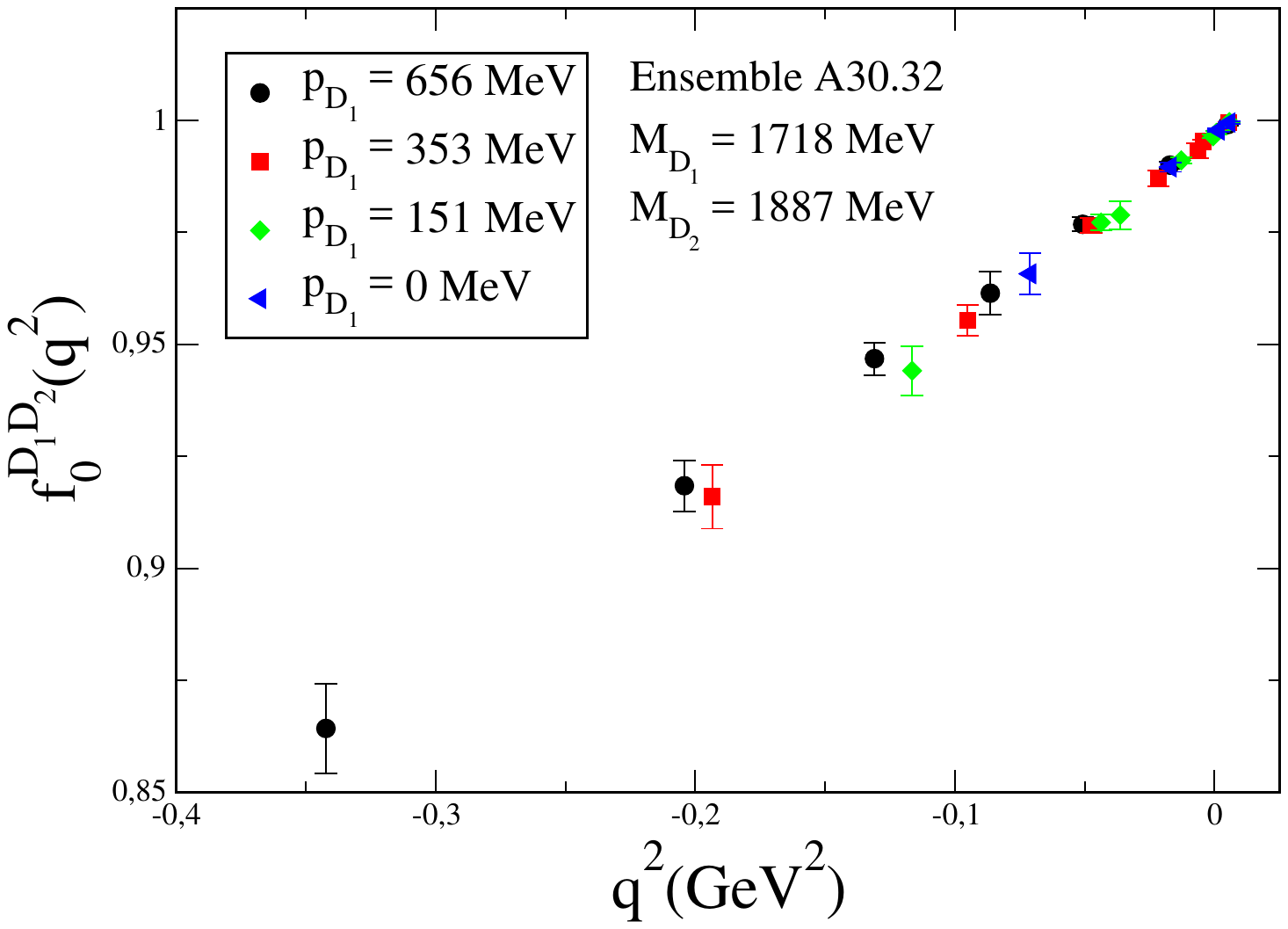}
}
\caption{\it Momentum dependence of the vector (left panel) and scalar (right panel) form factors regulating the semileptonic decay in which the parent and child mesons are two charmed PS mesons, $D_1$ and $D_2$, with masses close to the $D$-meson one. In this plot we have used the gauge ensemble A30.32, in which $D_1$ and $D_2$ have masses equal to $1718$ MeV and $1887$ MeV, respectively. Different markers and colors distinguish different values of the child meson momentum.}
\label{fig:DtoD}
\end{figure}

Before closing the Section, it is worth noting that no evidence of hypercubic effects within the current statistical uncertainties was found in the case of the $K \to \pi \ell \nu$ semileptonic form factors analyzed in Ref.~\cite{Carrasco:2016kpy}, where the same gauge configurations and the same parent and child momenta were adopted\footnote{In Ref.~\cite{Carrasco:2016kpy} the vector and scalar form factors for the $K_{\ell 3}$ decays have been constructed using local interpolating fields for both the pion and the kaon. We have checked that no hypercubic effects are visible also in the case of smeared interpolating fields.}. 
This suggests that the hypercubic artifacts may be governed by the difference between the parent and the child meson masses. 
Such an indication is confirmed by the results given in Fig.~\ref{fig:DtoD}, where the transition between two charmed PS mesons with masses close to the $D$-meson one has been considered. 
The momentum dependencies of the corresponding form factors show no evidence of hypercubic effects within the statistical uncertainties.
The dependence of hypercubic artifacts upon the mass difference between the parent and the child mesons is clearly a very important issue, which warrants further investigations. 
It may represent an important warning in the case of the determination of the form factors governing semileptonic $B$-meson decays into lighter mesons.

\section{Subtraction of the hypercubic effects}
\label{sec:sec3}

As shown in the previous Section the form factors $f_+$ and $f_0$ for both the $D \to \pi$ and $D \to K$ decays exhibit a sizeable Lorentz-symmetry breaking due to hypercubic effects generated at finite lattice spacing. 
Let's start by introducing Euclidean 4-momenta defined, in the case of the 4-momentum transfer $q_\mu = (q_0, ~ \vec{q})$, as
 \be
      q_\mu^E = \left( \vec{q}, ~ q_4 \right) = \left( \vec{q}, ~ iq_0 \right)
      \label{eq:Euclidean}
 \ee
so that $\sum_\mu q_\mu^E q_\mu^E = - q^2$.
Hypercubic invariants can be constructed using the two momenta $q_\mu^E$ and $P_\mu^E = (p_D + p_P)_\mu^E$ as
 \be
      q^{[n]} P^{[m]} \equiv \sum_\mu \left( q_\mu^E \right)^n \left( P_\mu^E \right)^m ~ .
      \label{eq:hypercubic}
 \ee
For $n + m  = 2$ there are three invariants, which are Lorentz invariants, namely $q^2$, $q \cdot P$ and $P^2$. 
The three invariants can be rewritten in terms of $q^2$ and the parent and child meson masses.
For $n + m = 4$ there are five hypercubic invariants: $q^{[4]}$, $q^{[3]} P^{[1]}$, $q^{[2]} P^{[2]}$, $q^{[1]} P^{[3]}$, $P^{[4]}$, where $q^{[4]}$ ($P^{[4]}$) stands for $q^{[4]} P^{[0]}$ ($q^{[0]} P^{[4]}$).

As already shown in Fig.~\ref{fishbone}, the form factors $f_{+, 0}$ calculated using Eqs.~(\ref{eq:V0_final}-\ref{eq:S_final}) do not depend on $q^2$ only.
A possible way to describe the observed hypercubic effects is to address them directly on the vector and scalar matrix elements.
We start by considering the following decomposition of the vector matrix elements:
 \be
    \braket{P(p_P) | \widehat{V}_\mu^E | D(p_D)} = \braket{\widehat{V}_\mu^E}_{\rm Lor} + \braket{\widehat{V}_\mu^E}_{\rm hyp} ~ ,
    \label{eq:vector_decomposition}
 \ee
where $\braket{\widehat{V}_\mu^E}_{\rm Lor}$ is the Lorentz-covariant term
 \be
    \braket{\widehat{V}_\mu^E}_{\rm Lor} = P_\mu^E~ f_+(q^2, a^2) + q_\mu^E ~ \frac{M_D^2 - M_P^2}{q^2} \left[ f_0(q^2, a^2) - f_+(q^2, a^2) \right] ~ ,
    \label{eq:vector_Lorentz}
 \ee
while $\braket{\widehat{V}_\mu^E}_{\rm hyp}$ is given by
 \be
    \braket{\widehat{V}_\mu^E}_{\rm hyp} = a^2 \left[ \left( q_\mu^E \right)^3 ~ H_1+ \left( q_\mu^E \right)^2 P_\mu^E ~ H_2 + 
                                                                    q_\mu^E \left( P_\mu^E \right)^2 ~ H_3 + \left( P_\mu^E \right)^3 ~ H_4 \right]
    \label{eq:vector_hypercubic}
 \ee
and the quantities $H_i$ ($i = 1, 2, 3, 4$) are additional hypercubic form factors.
Note that in the Lorentz-covariant term (\ref{eq:vector_Lorentz}) we have explicitly considered that the form factors $f_{+,0}$ can be affected by discretization errors of order $\mathcal{O}(a^2)$, which are unrelated to hypercubic effects and may depend on $q^2$ as well as on the parent and child meson masses.

Eq.~(\ref{eq:vector_hypercubic}) is the most general structure, up to order $\mathcal{O}(a^2)$, that transforms properly under hypercubic rotations and is built with third powers of the components of the two momenta $q_\mu^E$ and $P_\mu^E$. 
The Lorentz-invariance breaking effects are encoded in the four structures $\left( q_\mu^E \right)^3$, $ \left( q_\mu^E \right)^2 P_\mu^E$, $q_\mu^E \left( P_\mu^E \right)^2$ and $\left( P_\mu^E \right)^3$ as well as in the hypercubic form factors $H_i$, which, we assume, depend only on $q^2$ and the parent and child meson masses. 
Note that the decomposition (\ref{eq:vector_decomposition}-\ref{eq:vector_hypercubic}) implies that the form factors $f_{+,0}$ calculated using Eqs.~(\ref{eq:V0_final}-\ref{eq:S_final}) do depend not only on $q^2$, but also on the five hypercubic invariants $q^{[4]}$, $q^{[3]} P^{[1]}$, $q^{[2]} P^{[2]}$, $q^{[1]} P^{[3]}$, $P^{[4]}$.

For the $H_i$ form factors we adopt a simple polynomial form in terms of the $z$ variable \cite{Boyd:1995cf,Arnesen:2005ez}
 \be
    H_i(z) = d_0^i + d_1^i z + d_2^i z^2 ~ ,
    \label{eq:Hi}
 \ee
where $z$ is defined as 
 \be
    z = \frac{{\sqrt {t_ +   - q^2 }  - \sqrt {t_ +   - t_0 } }}{{\sqrt {t_ +   - q^2 }  + \sqrt {t_ +   - t_0 } }}
 \ee
with $t_+$ and $t_0$ given by
 \bea
        t_+  & = & \left( M_D  + M_P \right)^2 ~ , \nonumber \\
        t_0  & = & \left( M_D  + M_P \right) \left( \sqrt {M_D}  - \sqrt {M_P} \right)^2 ~ .
\eea
In Eq.~(\ref{eq:Hi}) the coefficients $d_{0,1,2}^i$ will be treated as free parameters. 

Let's now turn to the matrix elements of the scalar density.
We consider the following decomposition of the scalar density matrix elements:
 \be
    \braket{P(p_P) | S | D(p_D)} = \braket{S}_{\rm Lor} + \braket{S}_{\rm hyp} ~ ,
    \label{eq:scalar_decomposition}
 \ee
where $\braket{S}_{\rm Lor}$ is the Lorentz-invariant term
 \be
    \braket{S}_{\rm Lor} = \frac{M_D^2 - M_P^2}{\mu_c - \mu_q} f_0(q^2, a^2) ~ ,
    \label{eq:scalar_Lorentz}
 \ee
while $\braket{S}_{\rm hyp}$ is given by
 \be
    \braket{S}_{\rm hyp} = \frac{a^2}{\mu_c - \mu_q} \left[ q^{[4]} ~ \widetilde{H}_1+ q^{[3]} P^{[1]} ~ \widetilde{H}_2 + 
                                        q^{[2]} P^{[2]} ~ \widetilde{H}_3 + q^{[1]} P^{[3]} ~ \widetilde{H}_4 + P^{[4]} \widetilde{H}_5\right]
    \label{eq:scalar_hypercubic}
 \ee
and the quantities $\widetilde{H}_i$ ($i = 1, 2, 3, 4, 5$) are additional hypercubic form factors.

\begin{figure}[htb!]
\centering{
\includegraphics[scale=0.30]{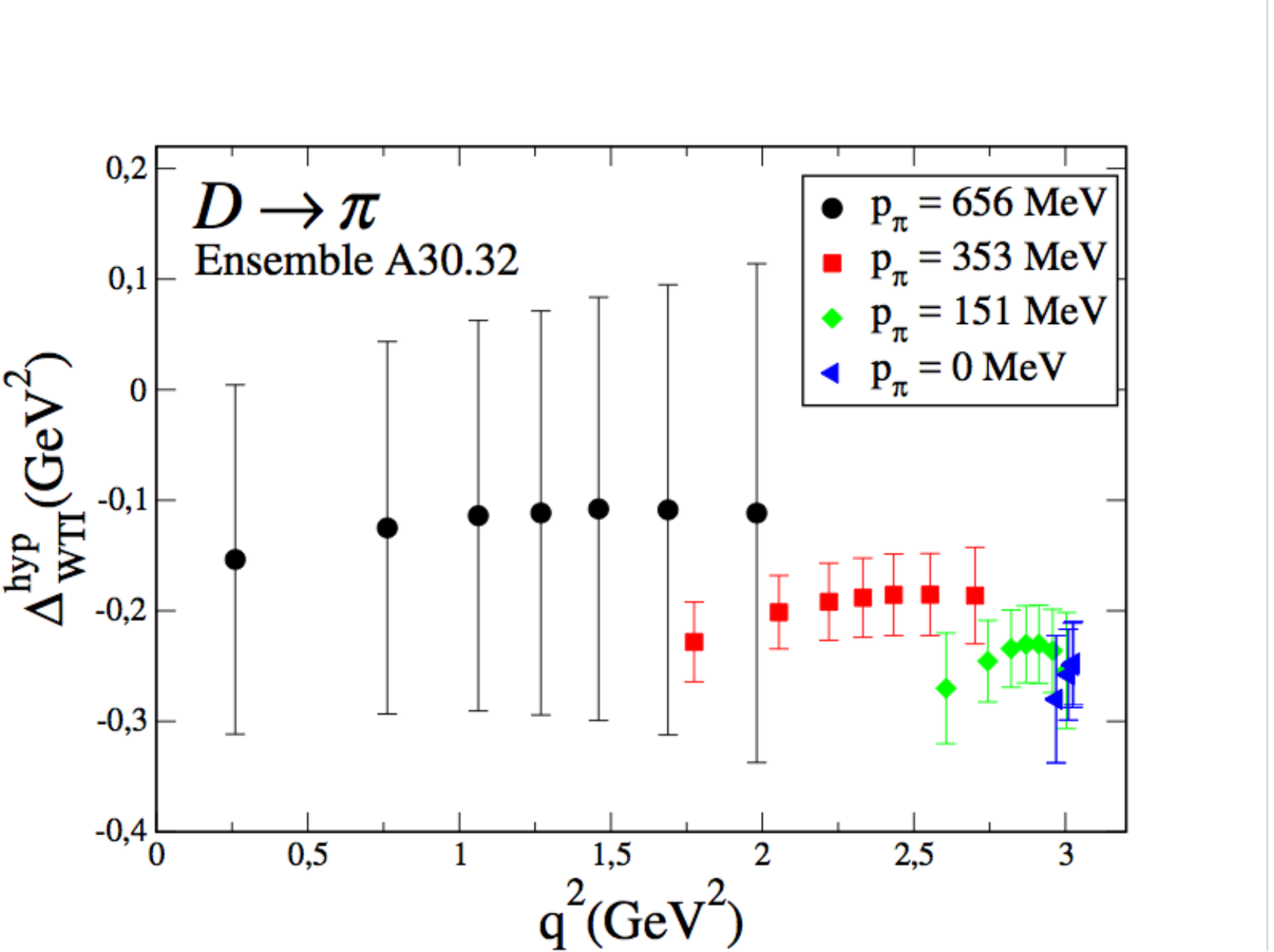} ~ 
\includegraphics[scale=0.30]{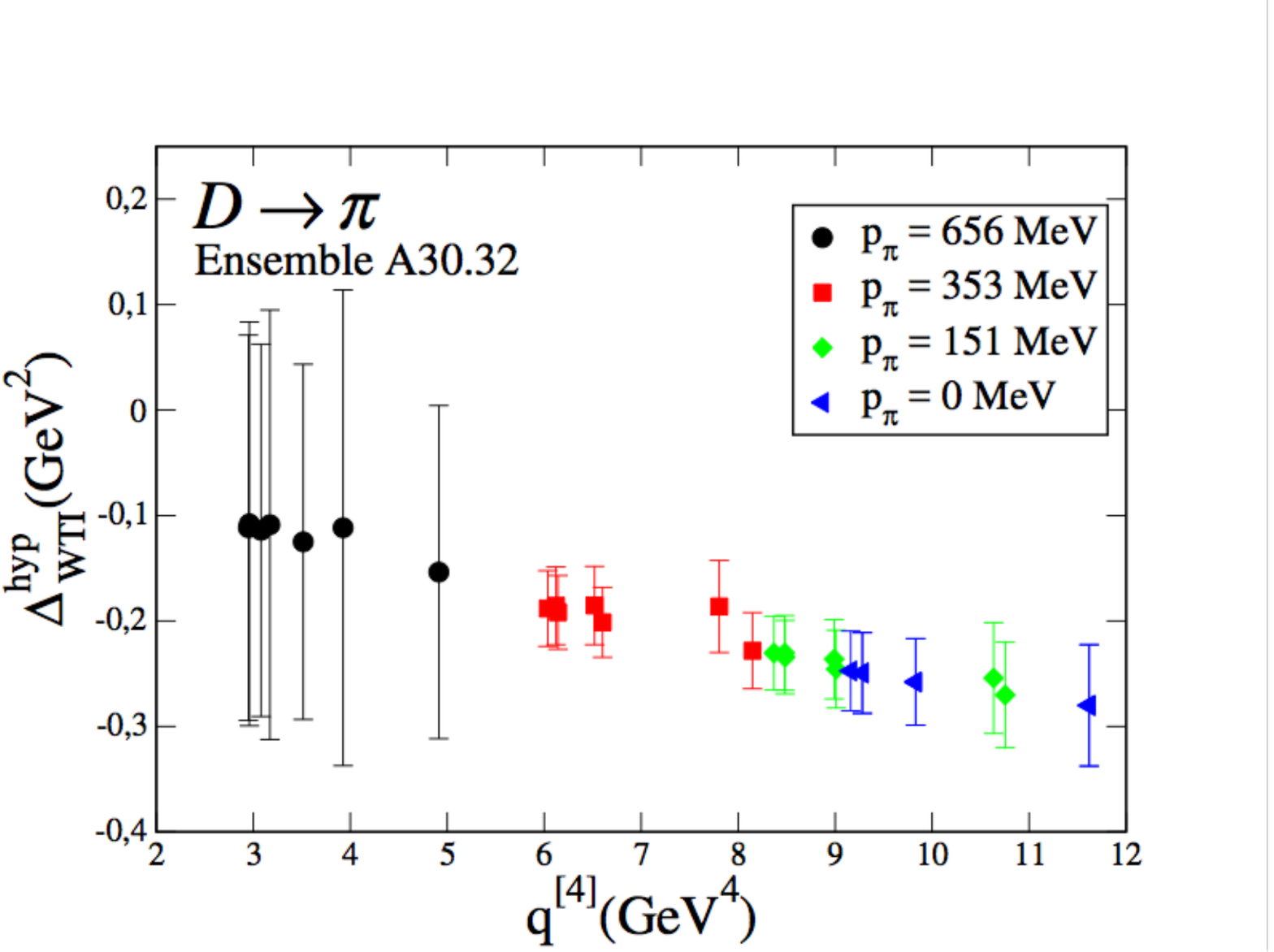}
\includegraphics[scale=0.30]{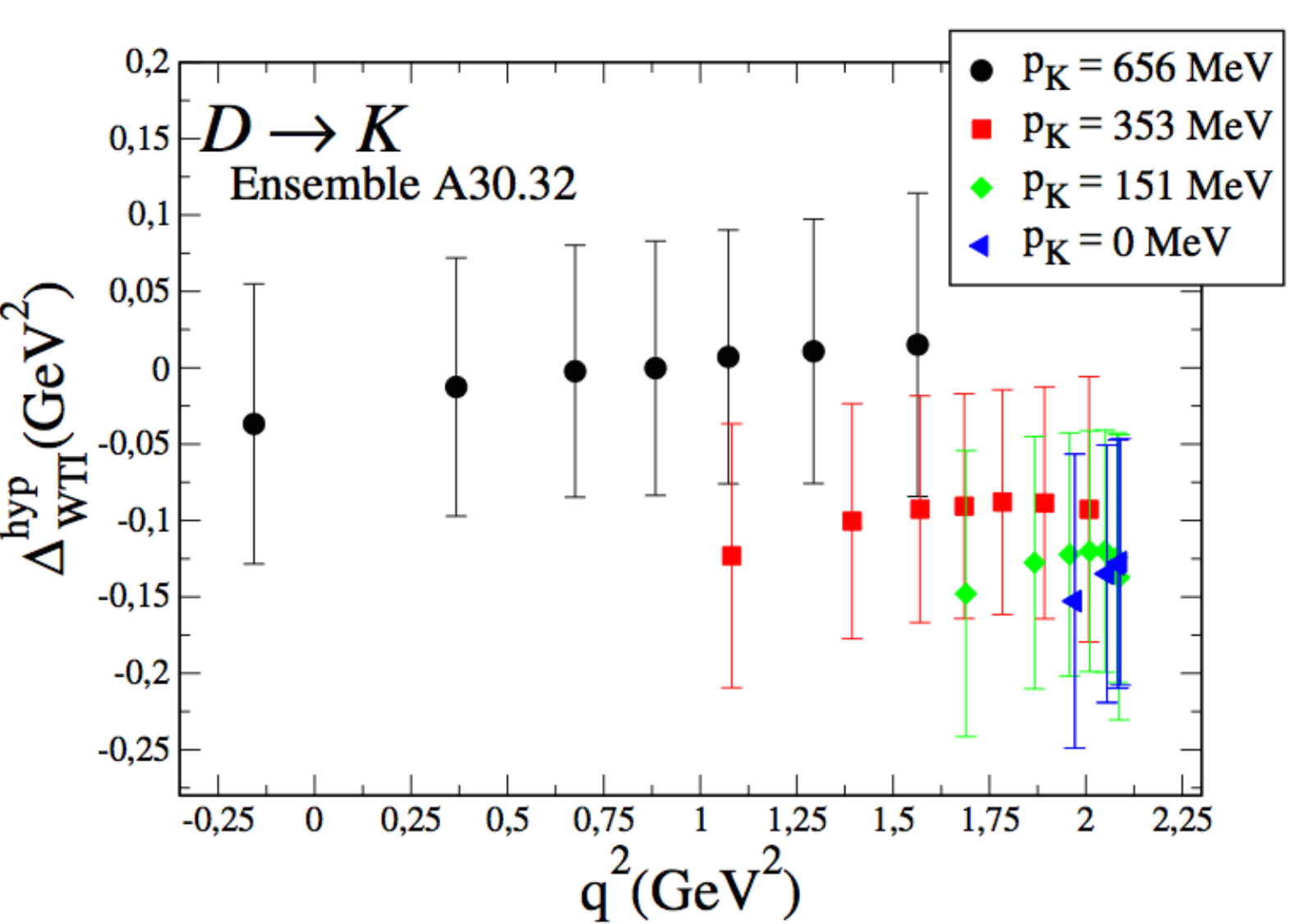} ~ 
\includegraphics[scale=0.30]{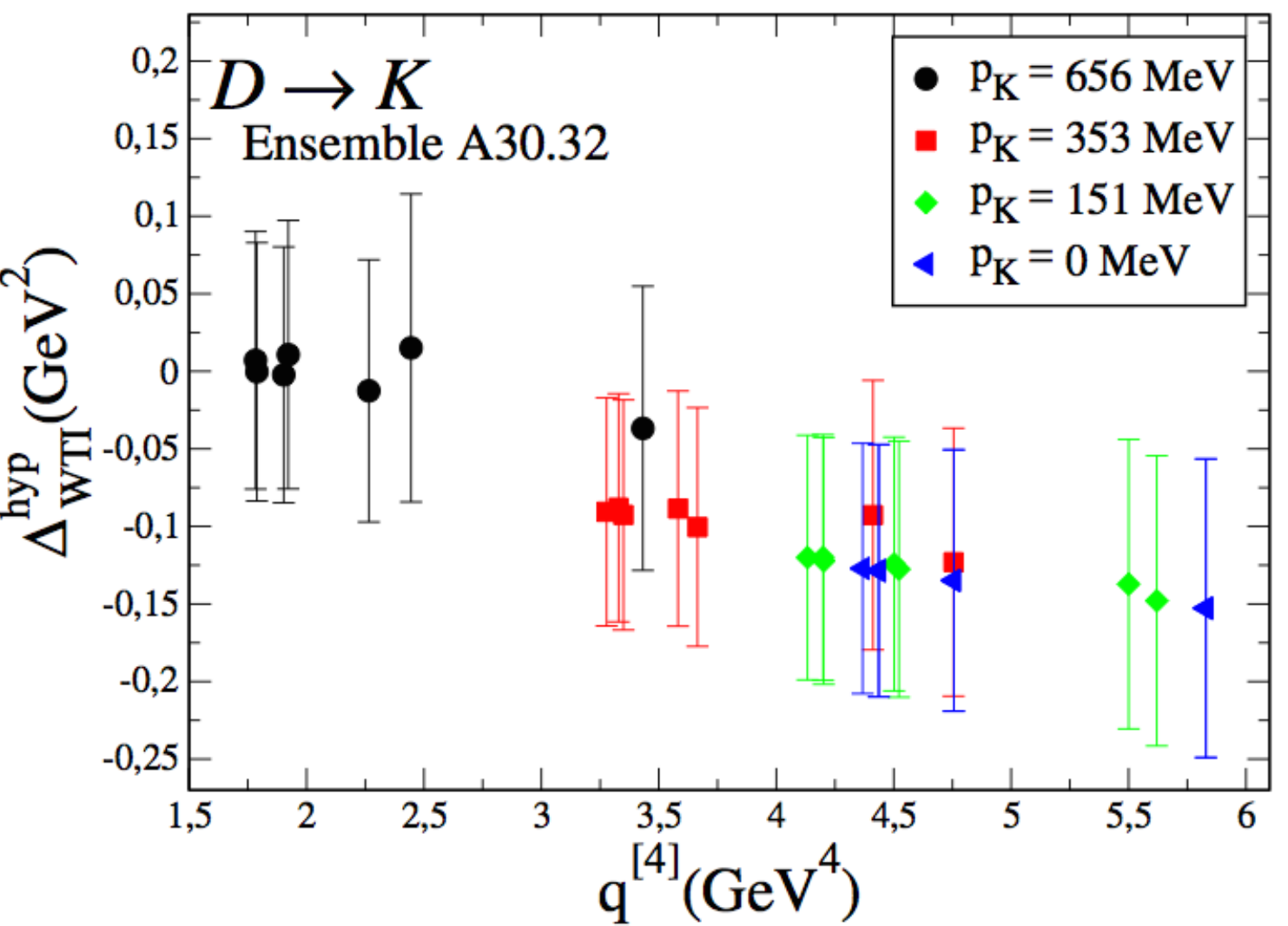}
}
\caption{\it Results for $\Delta_{WTI}^{\rm hyp}$ (see Eq.~(\protect\ref{eq:WTI})) versus $q^2$ (left panels) and $q^{[4]}$ (right panels) for the $D \to \pi$ (upper panels) and $D \to K$ (lower panels) transitions in the case of the gauge ensemble A30.32.}
\label{fig:WTI_violation}
\end{figure}

The Ward-Takahashi Identity (WTI) relates the 4-divergence of the vector current to the scalar density.
Let's introduce the WTI breaking term $ \Delta_{WTI}^{\rm hyp}$ defined as
 \bea
     \Delta_{WTI}^{\rm hyp} & = & \left(\mu_c - \mu_q \right) \braket{P(p_P) | S | D(p_D)} + q_\mu^E \braket{P(p_P) | \widehat{V}_\mu^E | D(p_D)} 
                                                     \nonumber \\
                                           & = & \left(\mu_c - \mu_q \right) \braket{S}_{\rm hyp} + q_\mu^E \braket{\widehat{V}_\mu^E}_{\rm hyp}
     \label{eq:WTI}
 \eea
which implies
 \bea
     \Delta_{WTI}^{\rm hyp} & = & a^2 \left[ q^{[4]} ~ (\widetilde{H}_1 + H_1)+ q^{[3]} P^{[1]} ~ (\widetilde{H}_2 + H_2) + q^{[2]} P^{[2]} ~ (\widetilde{H}_3 + H_3) 
                                                     \right. \nonumber \\
                                           & + & \left. q^{[1]} P^{[3]} ~ (\widetilde{H}_4 + H_4) + P^{[4]} \widetilde{H}_5\right] ~ .
     \label{eq:WTI_general}
 \eea
The quantity $\Delta_{WTI}^{\rm hyp}$ can be evaluated directly using the matrix elements $\braket{\widehat{V}_\mu^E}$ and $\braket{S}$.
Its dependence on the parent child momenta is illustrated in Fig.~\ref{fig:WTI_violation} in the case of the gauge ensemble A30.32.
It can be seen that the WTI-violating term $\Delta_{WTI}^{\rm hyp}$ is small, but not vanishing and it cannot depend only on the Lorentz-invariant $q^2$.
Instead the lattice data suggest a simple, approximate linear dependence on the hypercubic invariant $q^{[4]}$ (see right panels in Fig.~\ref{fig:WTI_violation}).
This implies a quite simplified structure for $\Delta_{WTI}^{\rm hyp}$ in Eq.~(\ref{eq:WTI_general}) and consequently for $\braket{S}^{\rm hyp}$ in Eq.~(\ref{eq:scalar_hypercubic}), namely: $\widetilde{H}_i + H_i = 0$ for $i = 2, 3, 4$, $\widetilde{H}_5 = 0$ and $\widetilde{H}_1 + H_1 = H_S \neq 0$.
In other words one has $\Delta_{WTI}^{\rm hyp} = a^2 q^{[4]} H_S$ and 
 \be
    \braket{S}_{\rm hyp} = \frac{a^2}{\mu_c - \mu_q} q^{[4]} ~ H_S - \frac{1}{\mu_c - \mu_q} q_\mu^E \braket{\widehat{V}_\mu^E}_{\rm hyp} ~ ,
    \label{eq:scalar_hypercubic_final}
 \ee
where for the hypercubic form factor $H_S$ we adopt the simple Ansatz 
 \be
      H_S = d_0^S + d_1^S m_\ell
      \label{eq:HS}
 \ee
with $d_{0,1}^S$ being free parameters.

The structure of the hypercubic artifacts is thus given by Eqs.~(\ref{eq:vector_hypercubic}) and (\ref{eq:scalar_hypercubic_final}) in terms of five hypercubic form factors $H_i$ ($i = 1, 2, 3, 4, S$).
The latter cannot be determined by analyzing the matrix elements of vector and scalar currents calculated separately for each gauge ensemble in the present work.
A simultaneous, global fit of all the data (more than one thousand lattice points corresponding to the time and spatial components of the vector and scalar currents for the 15 ETMC gauge ensembles of Table~\ref{tab:simu&masses}) has to be performed by considering the dependencies on $q^2$, $m_\ell$ and $a^2$ of the form factors $f_{+,0}$ as well as the $q^2$ and $m_\ell$ dependencies of the five hypercubic form factors $H_i$ ($i=1, 2, 3, 4, S$). 

For the form factors $f_{+,0}(q^2, a^2)$ we have adopted the modified z-expansion of Ref.~\cite{Bourrely:2008za}, viz.
 \bea
   \label{eq:z-exp_f+}
    f_+^{D \to \pi(K)}(q^2, a^2) \!\!\!& = &\!\!\! \frac{f^{D \to \pi(K)}(0, a^2) + c^{D \to \pi(K)}_+(a^2)\, (z - z_0)
                                                               \left(1 + \frac{z + z_0}{2} \right)}{\mathcal{P}^{D \to \pi(K)}_+(q^2)} ~ ,\\
   \label{eq:z-exp_f0}
    f_0^{D \to \pi(K)}(q^2, a^2) \!\!\!& = &\!\!\! \frac{f^{D \to \pi(K)}(0, a^2) + c^{D \to \pi(K)}_0(a^2)\, (z - z_0) 
                                                               \left(1 + \frac{z + z_0}{2} \right)}{\mathcal{P}^{D\to\pi(K)}_0(q^2)} ~ ,
 \eea
where we assume for the coefficients $c_{+(0)}^{D \to \pi(K)}(a^2)$ a simple linear dependence on $a^2$ and $z_0 \equiv z(q^2 = 0$), so that the condition $f_+(0, a^2) = f_0(0, a^2) = f(0, a^2)$ is explicitly fulfilled at finite lattice spacing.
In the r.h.s.~of Eqs.~(\ref{eq:z-exp_f+}-\ref{eq:z-exp_f0}) the terms at second order in the $z$-variable are constrained by the analyticity requirements described in Ref.~\cite{Bourrely:2008za}.

As for the functions $\mathcal{P}_{+,0}$, in the case of $D \to \pi$ transition we adopt the single-pole expressions
 \bea
   \label{eq:pole_f+_pi}
    \mathcal{P}_+^{D\to\pi}(q^2) \!\!\!&=&\!\!\! 1 - \frac{q^2}{M_V^2} ~,\\
   \label{eq:pole_f0_pi}
    \mathcal{P}_0^{D\to\pi}(q^2) \!\!\!&=&\!\!\! 1 - K_{FSE}^0(L)\,\frac{q^2}{M_S^2} ~,
 \eea
while for the $D \to K$ channel we use
 \bea
    \label{eq:pole_f+_K}
    \mathcal{P}_+^{D\to K}(q^2) \!\!\!&=&\!\!\!  1 - \frac{q^2}{M_{D_s^*}^2} \left( 1 + P_+ a^2 \right) ~,\\
    \label{eq:pole_f0_K}
    \mathcal{P}_0^{D\to K}(q^2) \!\!\!&=&\!\!\!  1 ~.
   \label{eq:poles}
 \eea
In the case of the $D \to \pi$ pole factors (\ref{eq:pole_f+_pi}) and (\ref{eq:pole_f0_pi}) the quantities $M_V$ and $M_S$ represent the vector and scalar pole masses, respectively. 
They are treated as free parameters in the fitting procedure.
In the case of the $D \to K$ decays the data are fitted equally well even excluding the pole term in the scalar form factor and therefore we choose $\mathcal{P}_0^{D \to K}(q^2) = 1$.
Conversely the physical vector meson $D_s^*$ has a mass below the cut threshold $\sqrt{t_+} = (M_{D_s} + M_K)$. 
Consequently the pole factor (\ref{eq:pole_f+_K}), including a simple discretization effect proportional to $a^2$, is introduced to guarantee the applicability of the $z$-expansion.

In Eq.~(\ref{eq:pole_f+_pi}) the quantity $K_{FSE}^0(L)$ takes into account the FSEs observed in Fig.~\ref{fig:FSE} by adopting the following phenomenological form
\be
   K_{FSE}^0(L) = 1 + C_{FSE}^{0} ~ \xi_\ell ~ \frac{e^{-M_\pi L}}{M_\pi L} ~,
   \label{eq:FSE}
\ee
where $C_{FSE}^0$ is a free parameter and $\xi_\ell = 2B m_\ell/ (16\pi^2f^2)$, with $B$ and $f$ being the SU(2) low-energy constants entering the LO chiral Lagrangian and determined in Ref.~\cite{Carrasco:2014cwa}. 

For the vector form factor at zero 4-momentum transfer, $f^{D \to \pi(K)}(0, a^2)$, we use the following Ansatz
\bea
    \label{eq:ChLim}
    f^{D \to \pi(K)}(0, a^2) \!\!\!&=&\!\!\! F_+ \left[ 1 + A^{\pi(K)}\, \xi_\ell \log\xi_\ell + b_1\, \xi_\ell + b_2 \, \xi^2_\ell + D \, a^2 \right] ~,
\eea
where the coefficients $F_+$, $b_1$, $b_2$ and $D$ are treated as free parameters in the fitting procedure, while $A^{\pi(K)}$ is the chiral-log coefficient predicted by the hard pion SU(2) Chiral Perturbation Theory (ChPT) \cite{Bijnens:2010jg}, given by
\bea
    \label{eq:HPChPT}
    A^\pi = - \frac{3}{4} \left( 1 + 3 \widehat{g}^2 \right) ~ , \qquad A^K= + \frac{1}{2} ~ ,
\eea
where for the coupling constant $\widehat{g}$ we adopt the value $\widehat{g} = 0.61$ \cite{Olive:2016xmw}.

Using the ingredients described above we have performed the global, combined fit of all the data 
for the matrix elements $\braket{\widehat{V}_0}$, $\braket{\widehat{V}_{sp}}$ and $\braket{S}$, which amount to a total of $1110$ data points for both the $D \to \pi$ and $D \to K$ transitions.
The total number of free parameters is $24$ ($19$) in the case of $D \to \pi(K)$ channel, namely:
\begin{itemize}
\item in Eq.~(\ref{eq:ChLim}) 3 parameters for $D \to \pi$ ($F_+$, $b_1$ and $D$) and 4 parameters for $D \to K$ ($F_+$, $b_1$, $b_2$ and $D$);

\item in Eq.~(\ref{eq:z-exp_f+}) 2 parameters for $c_+^{D \to \pi(K)}$  (i.e., $c_+ = c_+^0 + c_+^1 a^2$);

\item in Eq.~(\ref{eq:z-exp_f0}) 2 parameters for $c_0^{D \to \pi(K)}$ (i.e., $c_0 = c_0^0 + c_0^1 a^2$);

\item 1 parameter ($M_V$) in Eq.~(\ref{eq:pole_f+_pi}) for $D \to \pi$ and 1 parameter ($P_+$)  in Eq.~(\ref{eq:pole_f+_K}) for $D \to K$; 

\item 2 parameters in Eqs.~(\ref{eq:pole_f0_pi}) and (\ref{eq:FSE}) ($M_S$ and $C_{FSE}^0$) only for $D \to \pi$,;

\item 2 parameters in Eq.~(\ref{eq:HS}) for the hypercubic form factor $H_S$;

\item in Eq.~(\ref{eq:Hi}) 3 parameters for each of the four hypercubic form factors $H_1$, $H_2$, $H_3$ and $H_4$ for $D \to \pi$ and 2 parameters (i.e., $d_2^i = 0$ ) for $D \to K$.
  
\end{itemize}

The quality of the fit is quite good obtaining $\chi^2 / \rm{d.o.f.} \simeq 1.2$ for both the $D \to \pi$ and $D \to K$ transitions. 
We have tried to include extra terms in Eqs.~(\ref{eq:z-exp_f+}-\ref{eq:z-exp_f0}) either proportional to $z^2$ (including the analyticity requirement of Ref.~\cite{Bourrely:2008za} through an appropriate term proportional to $z^3$) or proportional to the light-quark mass $m_\ell$ in the coefficients $c_{+,0}^{D \to \pi(K)}$. 
Since the differences in the results for both the hypercubic corrections and the form factors $f_{+,0}$ are negligible with respect to the other errors and the values of the new parameters turn out to be consistent with $0$, such extended fits are not used for estimating systematic uncertainties.

From the global, combined fit we obtain both the momentum dependence of the Lorentz-invariant form factors $f_{+,0}$ and of the five hypercubic form factors $H_i$ ($i = 1, 2, 3, 4, S$). 
The momentum dependence of $f_{+,0}$, extrapolated to the physical pion mass and to the continuum and infinite volume limits, will be discussed and compared to the experimental data in Section~\ref{sec:sec4}. 
Here we compute the hypercubic form factors $H_i$ coming from the global fit in order to check the quality of the subtraction of the hypercubic effects for each gauge ensemble, i.e.~at finite lattice spacing and volume and for the unphysical pion masses given in Table~\ref{tab:simu&masses}.

\begin{figure}[htb!]
\centering{
\includegraphics[scale=0.30]{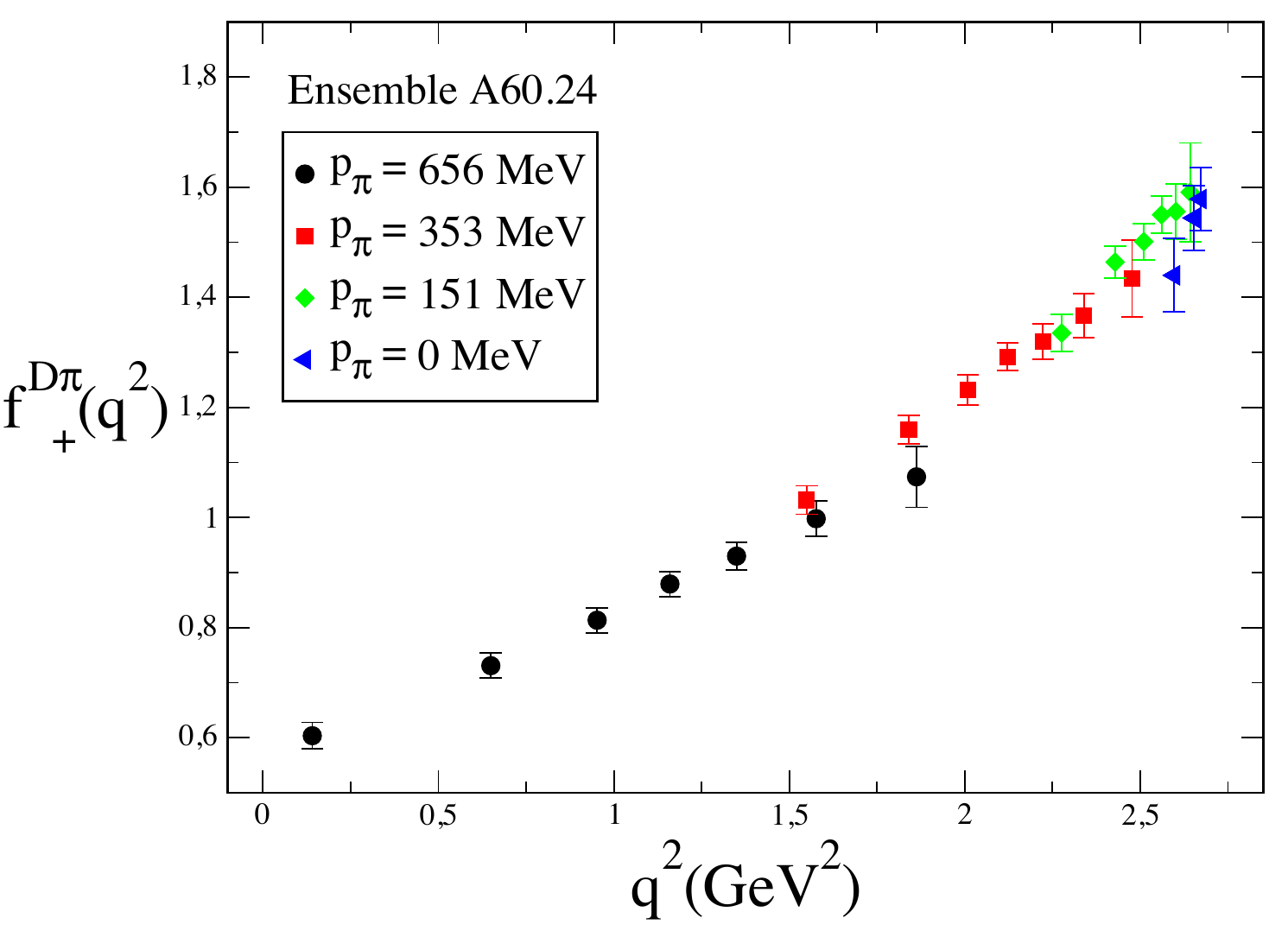} ~
\includegraphics[scale=0.30]{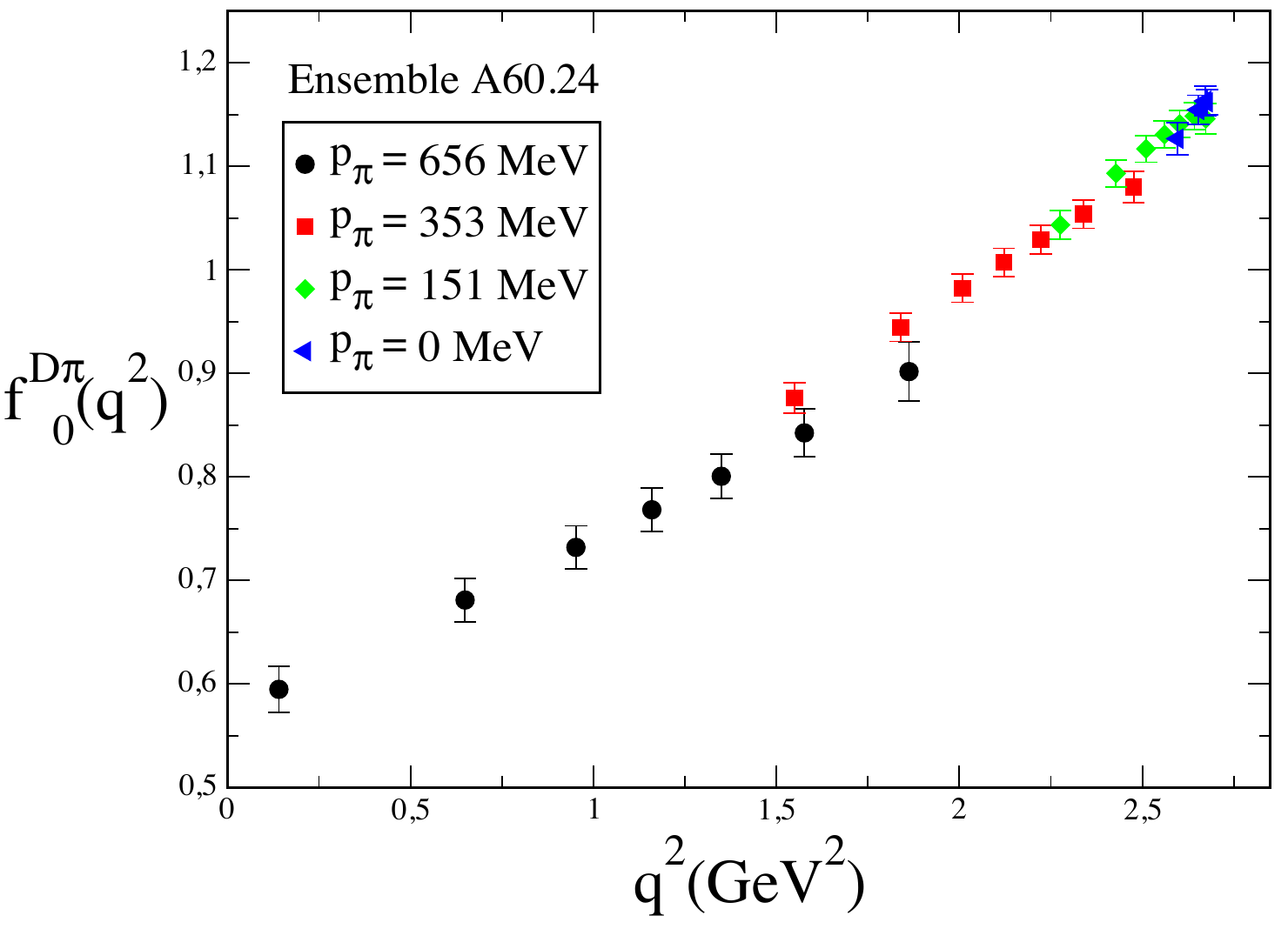}
\includegraphics[scale=0.30]{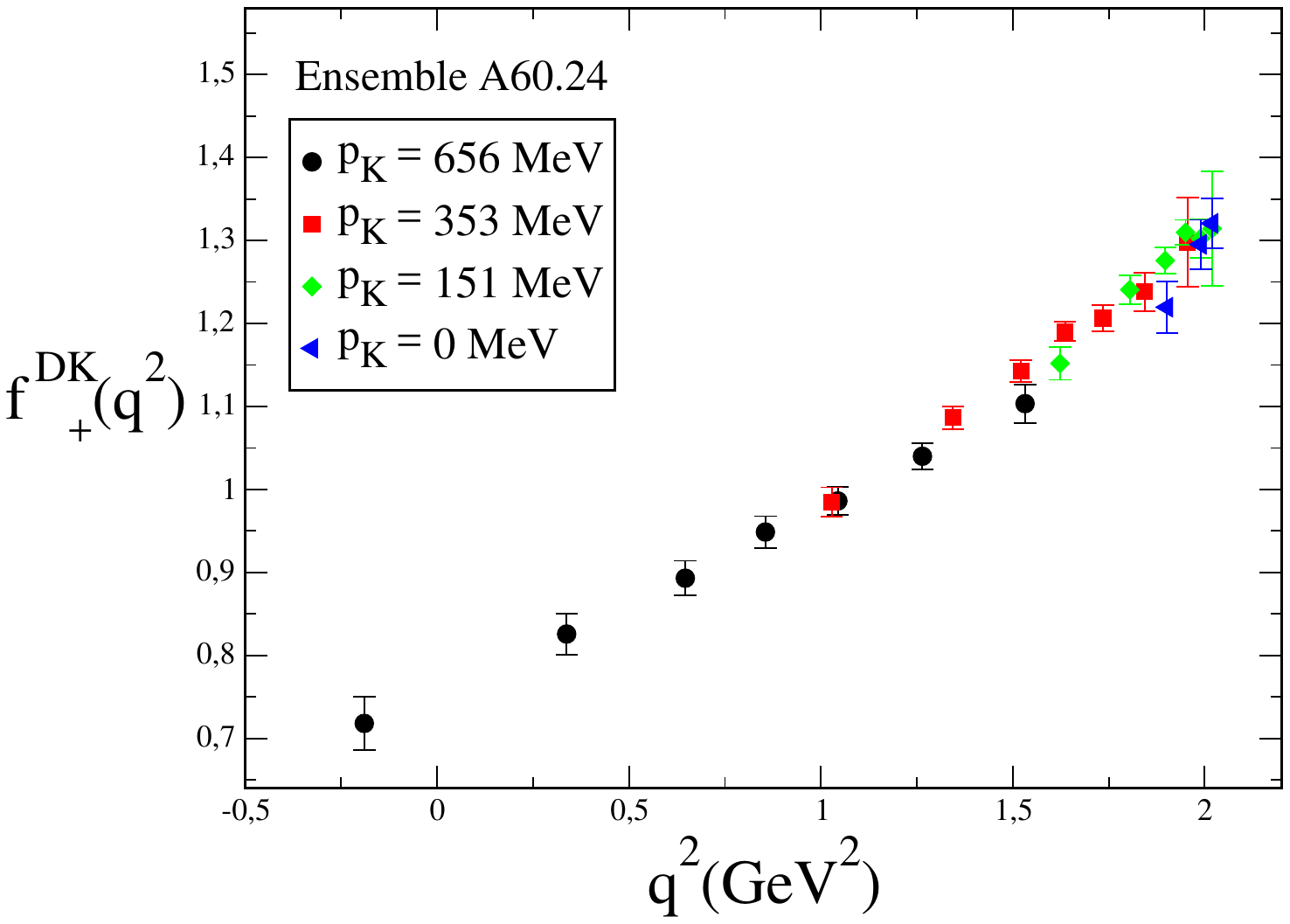} ~
\includegraphics[scale=0.30]{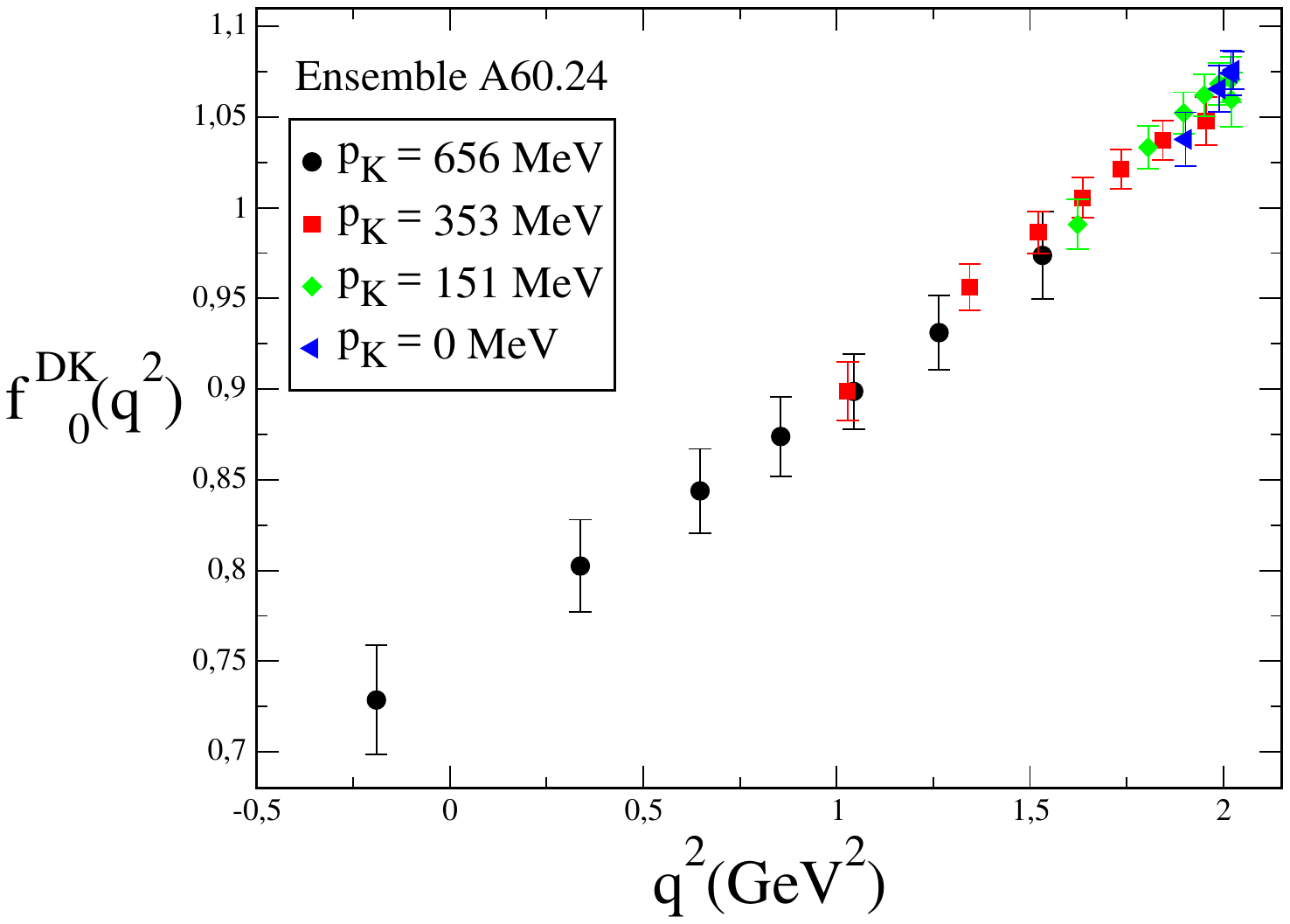}
}
\caption{\it The same as in Fig.~\ref{fishbone}, but after correcting for the hypercubic effects determined in the global fitting procedure using Eqs.~(\ref{eq:vector_decomposition}) and (\ref{eq:scalar_decomposition}). Different markers and colors distinguish different values of the child meson momentum.}
\label{fig:correction}
\end{figure}

In Fig.~\ref{fig:correction} we show the same form factors given in Fig.~\ref{fishbone} after the hypercubic contributions determined by the global fit have been subtracted from the matrix elements $\braket{\widehat{V}_\mu}$ and $\braket{S}$ using Eqs.~(\ref{eq:vector_decomposition}) and (\ref{eq:scalar_decomposition}). 
It can be seen that the hypercubic effects are properly removed and both the scalar and the vector form factors depend now only on the $4-$momentum transfer $q^2$ within the statistical uncertainties. 

In the limiting case where the parent and the child mesons are the same, Eq.~(\ref{eq:vector_hypercubic}) reduces to a simpler expression, namely
 \be
     \braket{D(p^\prime) | \widehat{V}_\mu^E |D(p)}_{\rm hyp} = a^2 \left[  \left( q_\mu^E \right)^2 P_\mu^E H_2 + \left( P_\mu^E \right)^3 H_4 \right] ~ ,
     \label{eq:vector_hypercubic_DD}
 \ee
because only even terms under the exchange of the initial and final PS mesons survive.
In Section~\ref{sec:sec2} we have noted that within the statistical uncertainties there is no evidence of hypercubic effects when the initial and final meson have the same masses  (see Fig.~\ref{fig:DtoD}).
This might be an indication that the hypercubic form factors $H_2$ and $H_4$ can be neglected. 
Thus, we have repeated the global fitting procedure assuming $H_2 = H_4 = 0$, which reduces
the number of free parameters to $18$ and $15$ for the $D \to \pi$ and $D \to K$ transitions, respectively. 

The differences in the results for both the hypercubic corrections and the form factors $f_{+,0}$, obtained including ($H_2 \neq H_4 \neq 0$) or excluding ($H_2 = H_4 = 0$) the two hypercubic form factors $H_2$ and $H_4$, are found to be negligible within the current statistical uncertainties.
Therefore, in what follows we adopt the fitting procedure, in which we assume $H_2 = H_4 = 0$, as our reference fit for estimating the uncertainties due to various sources of systematic errors as well as for  
obtaining the results for the form factors $f_{+,0}^{D \to \pi(K)}(q^2)$, extrapolated to the physical pion mass and to the continuum and infinite volume limits, which will be discussed and compared to the experimental data in Section~\ref{sec:sec4}.

We stress again that an important feature of our analysis with respect to previous studies of the semileptonic $D \to \pi(K)$ form factors is the use of a plenty of kinematical conditions corresponding to parent and child mesons either moving or at rest. 
Using only a limited number of kinematical conditions, for instance the Breit-frame in which $\vec{p_D} = - \vec{p}_{\pi(K)}$ or the $D-$meson at rest, the presence of the hypercubic effects may not be manifest.
\begin{figure}[htb!]
\centering{
\includegraphics[scale=0.30]{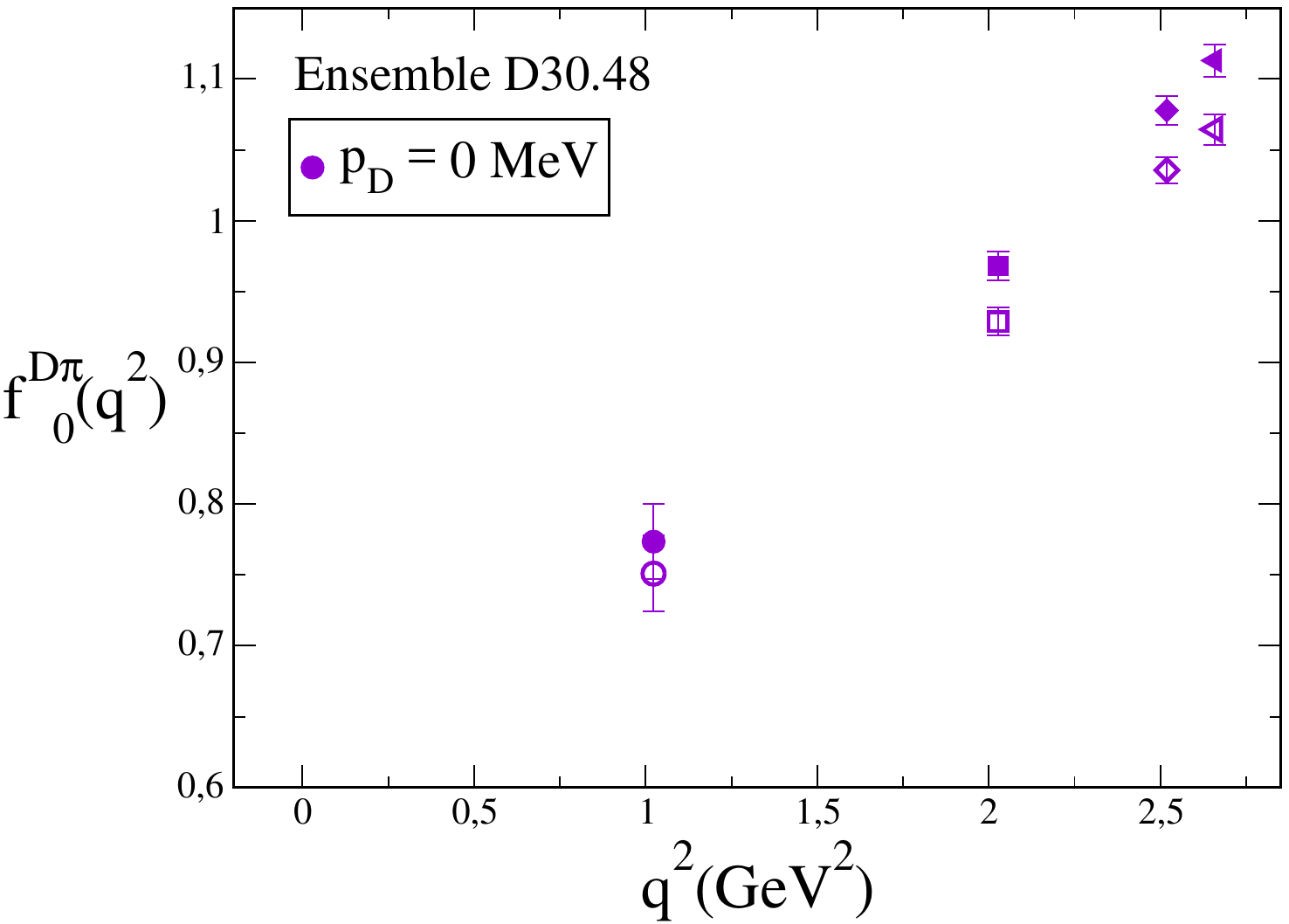} ~~ 
\includegraphics[scale=0.30]{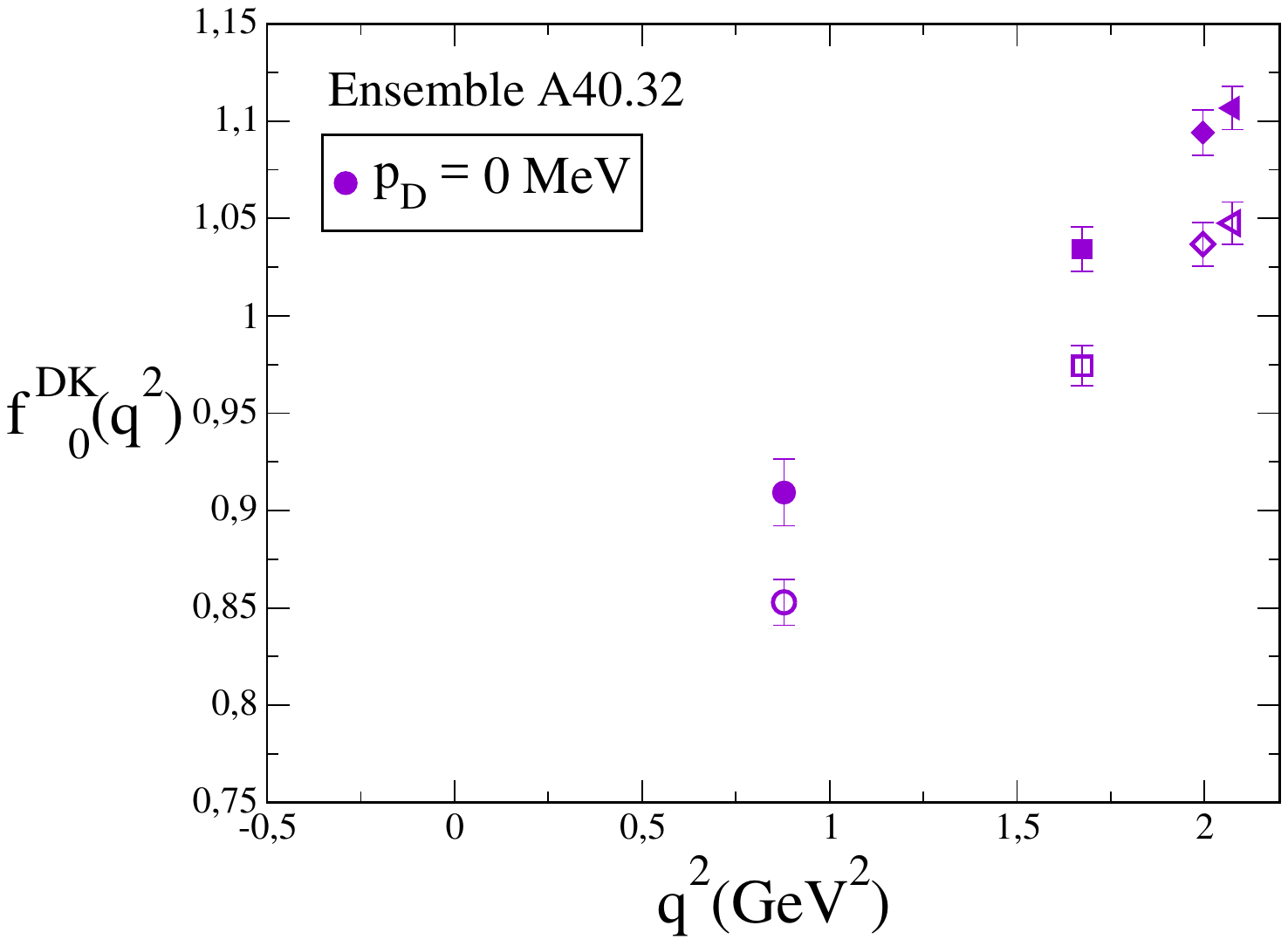}
}
\caption{\it Left panel: the scalar form factor $f_0^{D \pi}(q^2)$ corresponding to the kinematical conditions with the $D-$meson at rest for the gauge ensemble D30.48. Hollow and filled points represent, respectively, the data before and after the removal of the hypercubic effects determined in the global fitting procedure. Right panel: the same as in the left panel, but for $f_0^{D K}(q^2)$ in the case of ensemble A40.32}
\label{fig:D_at_rest}
\end{figure}
This point is illustrated in Fig.~\ref{fig:D_at_rest}, which shows the subset of our data for the scalar $D \to \pi$ (left panel) and $D\to K$ (right panel) form factor $f_0$ corresponding only to the D-meson at rest both before and after the subtraction of the hypercubic effects determined in the global fitting procedure.
Lorentz-symmetry breaking is not manifest in the limited set of data points with $\vec{p}_D = 0$, but it is not negligible.
This holds for the scalar form factor $f_0$, while in the case of the vector form factor $f_+$ we find that Lorentz-symmetry breaking effects are less pronounced in the subset of data corresponding to the D-meson at rest. 
We stress that the differences between the data with and without hypercubic effects are a ${\cal{O}}(a^2)$ effect proportional to hypercubic invariants. 
Thus, any analysis of the data without the subtraction of hypercubic effects, based directly on parameterizations like Eqs.~(\ref{eq:z-exp_f+}-\ref{eq:z-exp_f0}), where only discretization effects unrelated to hypercubic invariants are considered, is in principle inadequate and may lead to different results in the continuum limit.

\section{Results from the global fit and comparison with experimental data}
\label{sec:sec4}

\begin{figure}[t!]
\centering{
\includegraphics[scale=0.30]{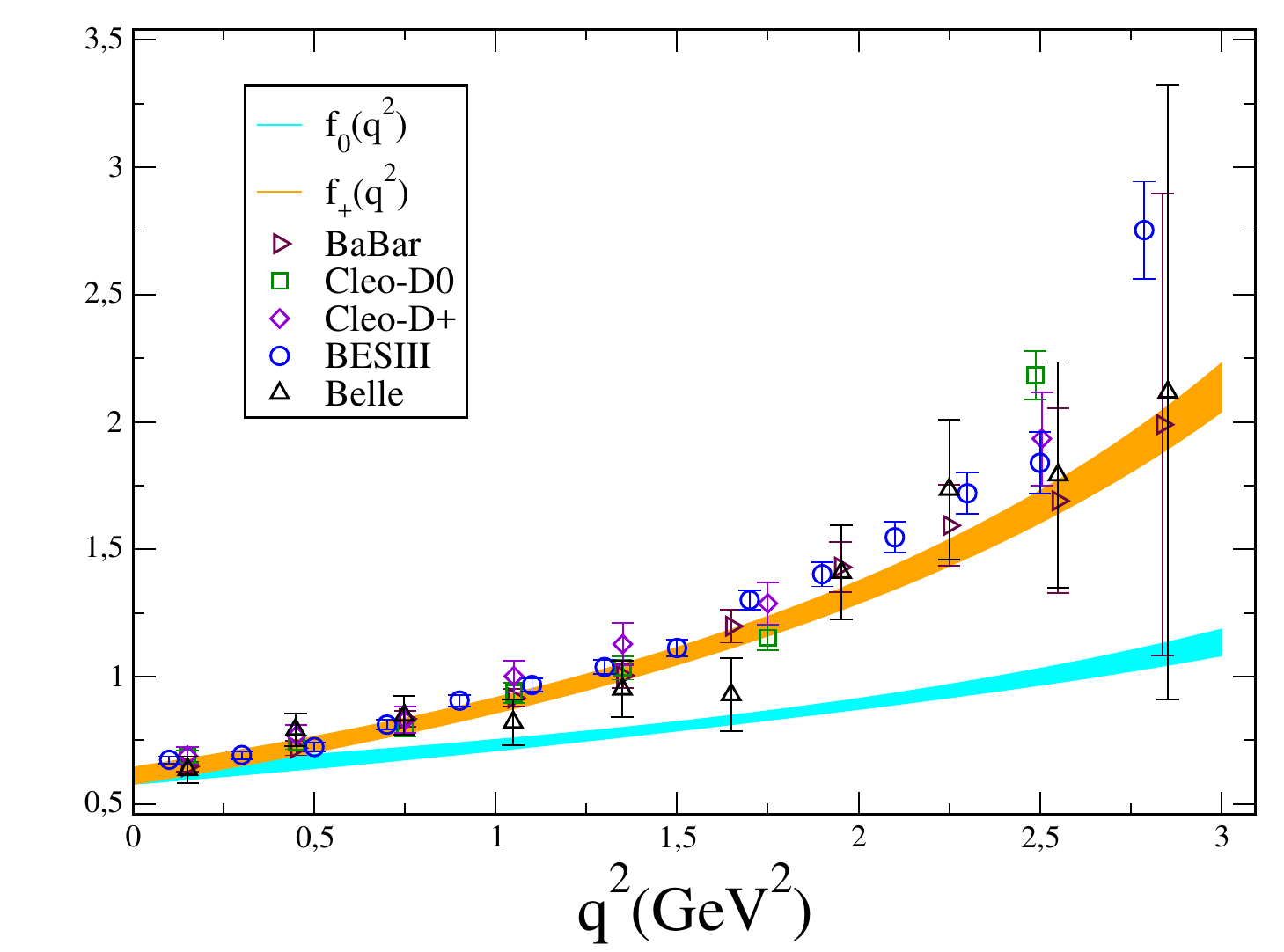} ~~ 
\includegraphics[scale=0.30]{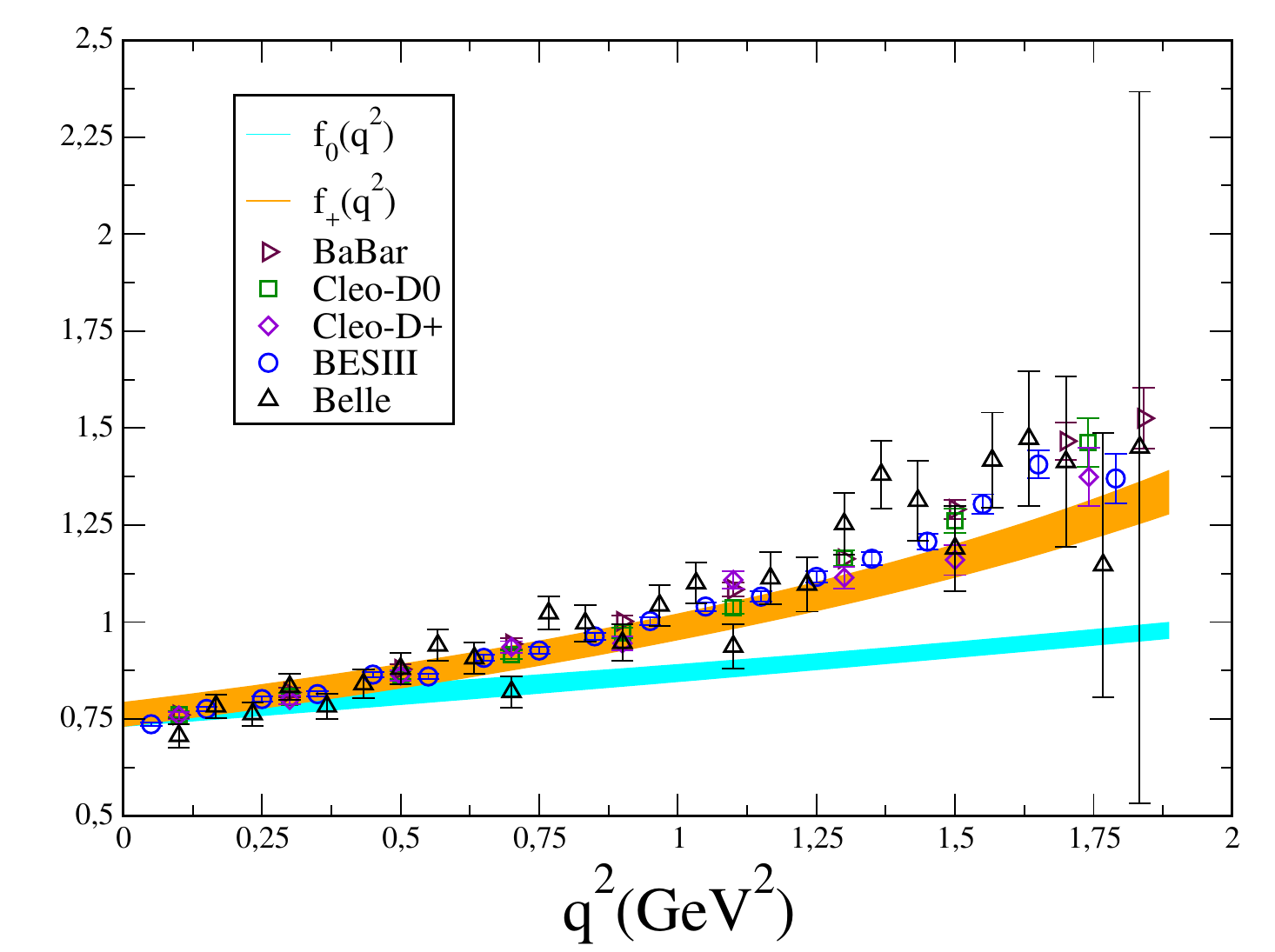}
}
\caption{\it Momentum dependencies of the Lorentz-invariant form factors $f_+(q^2)$ (orange bands) and $f_0(q^2)$ (cyan bands), extrapolated to the physical pion mass and to the continuum and infinite volume limits, for the $D \to \pi$ (left panel) and $D \to K$ (right panel) transitions, including their total uncertainties. For comparison, the values of $f^{D\pi(K)}_+(q^2)$ determined by BELLE, BABAR, CLEO and BESIII collaborations in Refs.~\cite{Widhalm:2006wz,Lees:2014ihu,Aubert:2007wg,Besson:2009uv,Ablikim:2015ixa} are shown. The bands correspond to the total (statistical + systematic) uncertainty at one standard-deviation level.}
\label{fig:physicalFormFactors}
\end{figure}

The momentum dependencies of the physical Lorentz-invariant vector and scalar form factors, extrapolated to the physical pion mass and to the continuum and infinite volume limits, are shown in Fig.~\ref{fig:physicalFormFactors} for both the $D \to \pi$ and $D \to K$ transitions.
Our results exhibit a remarkable precision in the full range of values of $q^2$ covered by the experiments (i.e., $0\leq q^2 \leq q^2_{\rm{max}} = (M_D - M_{\pi(K)})^2 \simeq 3.0 (1.9)$ GeV$^2$). 
Our results for the vector form factors $f^{D \pi}_+(q^2)$ and $f^{D K}_+(q^2)$ can be compared with the corresponding values determined by BELLE, BABAR, CLEO and BESIII collaborations in Refs.~\cite{Widhalm:2006wz,Lees:2014ihu,Aubert:2007wg,Besson:2009uv,Ablikim:2015ixa}, where the partial decay rates have been measured (see also Refs.~\cite{Rong:2014hea,Fang:2014sqa} for a summary of the experimental results).
The agreement is good except at high values of $q^2$, where some deviations are visible. 

\begin{figure}[htb!]
\centering{
\includegraphics[scale=0.30]{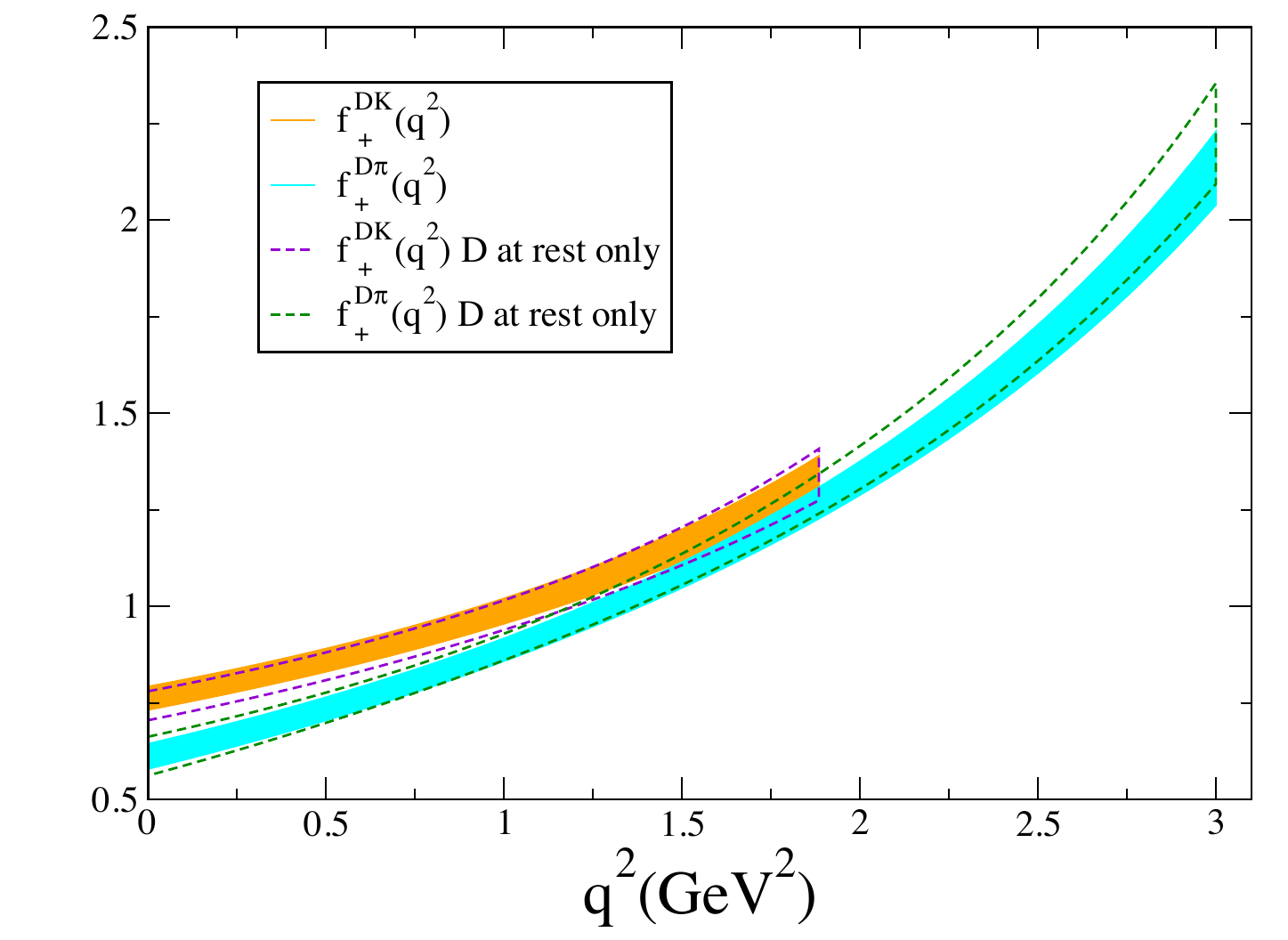} ~~  
\includegraphics[scale=0.30]{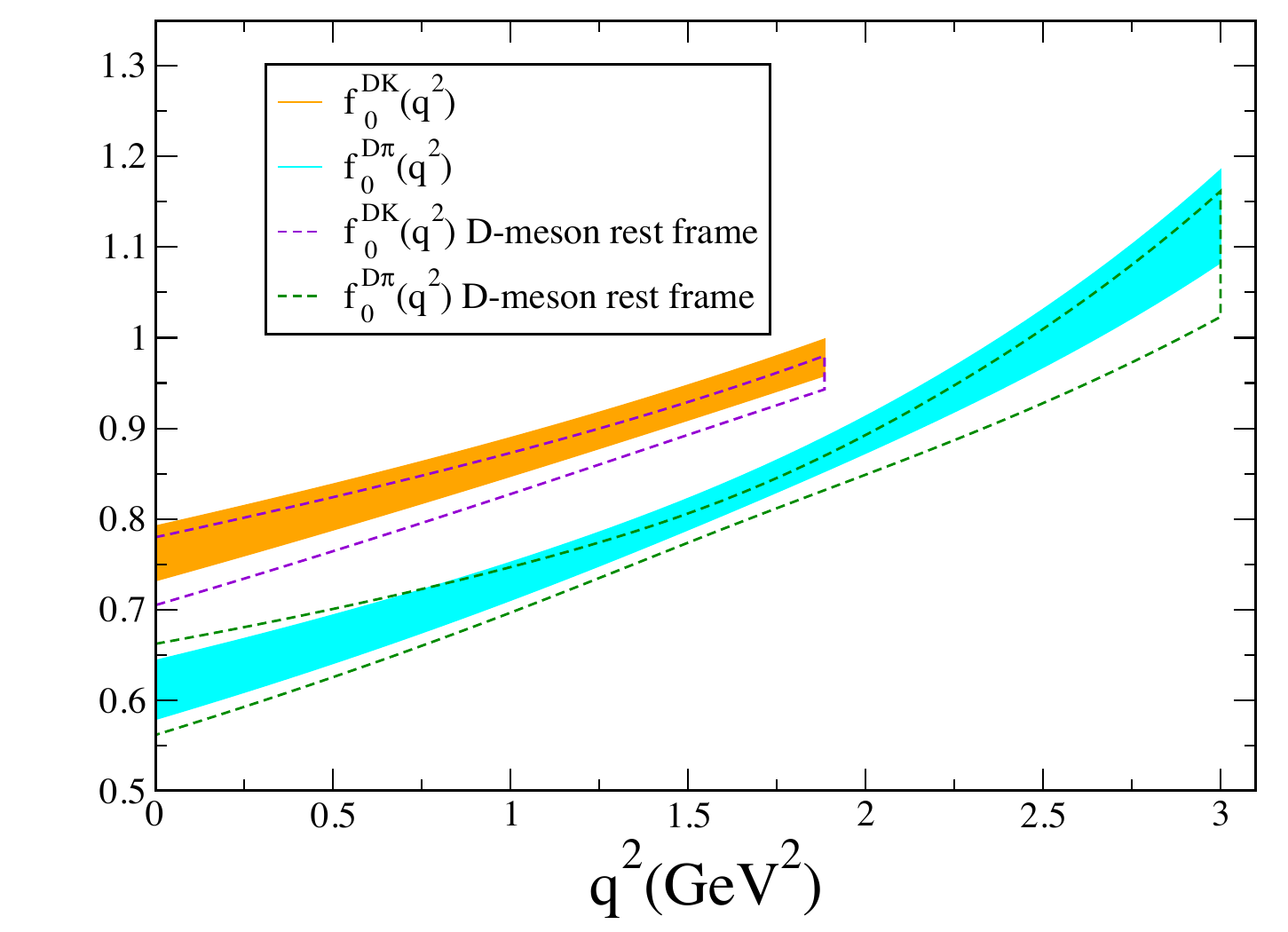}
}
\caption{\it Comparison of the vector and scalar form factors, extrapolated to the physical pion mass and to the continuum and infinite volume limits, obtained either by choosing all the kinematical configurations and including the hypercubic terms (\ref{eq:vector_hypercubic}) and (\ref{eq:scalar_hypercubic}) in the analysis (solid lines) or by limiting to the kinematical configurations corresponding to the $D$-meson rest frame without considering the subtraction of hypercubic effects (dashed lines). The bands correspond to the total uncertainty at one standard-deviation level.}
\label{fig:comparison_D_at_rest}
\end{figure}

In Fig.~\ref{fig:comparison_D_at_rest} our main results for the vector and scalar form factors are compared with those obtained by choosing only the kinematical configurations corresponding to the $D$-meson rest frame and by performing the extrapolations to the physical pion mass and to the continuum and infinite volume limits without including the hypercubic terms (\ref{eq:vector_hypercubic}) and (\ref{eq:scalar_hypercubic}). 
In other words, the continuum extrapolation is based only on the discretization terms contained in Eqs.~(\ref{eq:z-exp_f+}-\ref{eq:z-exp_f0}).
It can be seen that the neglect of hypercubic effects in the analysis and the use of a limited subset of data lead to some distortions of the extrapolated form factors, which are more pronounced in the case of the scalar form factor.
Such distortions are found to be comparable with present global uncertainties within one standard-deviation. 
They may become more relevant as the precision of the data will be increased in the future.  

In Table~\ref{tab:synthetic_Dpi} we provide a set of synthetic data points for the vector and scalar $D \to \pi$ form factors, $f^{D\pi}_+(q^2)$ and $f^{D\pi}_0(q^2)$, with the corresponding total uncertainties, calculated at eight selected values of $q^2$ between $0$ and $q^2_{\rm{max}} = (M_D - M_\pi)^2$. 
\begin{table}[htb!]
\renewcommand{\arraystretch}{1.2} 
\begin{center}
\begin{tabular}{|c|c|c|}
\hline
$q^2 ~ (\mbox{GeV}^2)$ & $f_+(q^2)$ & $f_0(q^2)$  \\ \hline 
 $0.0$    & ~ $0.612 ~ (35) ~~ (4) ~~ (7) ~ (1) ~ [35]$ ~ & ~ $0.612 ~ (35) ~~ (4) ~~ (7) ~ (1) ~ [35]$ ~  \\
 $0.4286$ & ~ $0.715 ~ (31) ~~ (4) ~~ (6) ~ (1) ~ [32]$ ~ & ~ $0.659 ~ (29) ~~ (4) ~~ (5) ~ (1) ~ [30]$ ~  \\
 $0.8571$ & ~ $0.840 ~ (29) ~~ (3) ~~ (6) ~ (1) ~ [30]$ ~ & ~ $0.713 ~ (24) ~~ (3) ~~ (3) ~ (1) ~ [24]$ ~  \\
 $1.2857$ & ~ $0.991 ~ (29) ~~ (4) ~~ (6) ~ (1) ~ [30]$ ~ & ~ $0.773 ~ (18) ~~ (4) ~~ (2) ~ (1) ~ [19]$ ~  \\
 $1.7143$ & ~ $1.179 ~ (34) ~ (10) ~~ (3) ~ (1) ~ [35]$ ~ & ~ $0.842 ~ (15) ~~ (4) ~~ (5) ~ (1) ~ [17]$ ~  \\
 $2.1429$ & ~ $1.415 ~ (43) ~ (15) ~~ (8) ~ (1) ~ [47]$ ~ & ~ $0.922 ~ (19) ~~ (5) ~~ (8) ~ (1) ~ [21]$ ~  \\ 
 $2.5714$ & ~ $1.721 ~ (60) ~ (21) ~ (16) ~ (1) ~ [66]$ ~ & ~ $1.017 ~ (29) ~~ (7) ~ (13) ~ (1) ~ [32]$ ~  \\  
 $3.0000$ & ~ $2.130 ~ (86) ~ (31) ~ (27) ~ (3) ~ [96]$ ~ & ~ $1.134 ~ (45) ~ (10) ~ (18) ~ (1) ~ [49]$ ~  \\ \hline
\end{tabular}
\end{center}
\renewcommand{\arraystretch}{1.0} 
\caption{\it Synthetic data points for the transition $D \to \pi$ representing our results for the vector and scalar form factors extrapolated to the physical pion point and to the continuum and infinite volume limits for eight selected values of $q^2$ in the range between $q^2 = 0$ and $q^2 = q_{max}^2 = (M_D - M_\pi)^2 \simeq 3.0 ~ \gev^2$. The errors correspond to the uncertainties related to (statistical + fitting procedure + input parameters), chiral extrapolation, FSEs and discretization effects, respectively (see text). The errors in squared brackets correspond to the combination in quadrature of the statistical and all systematic errors. }
\label{tab:synthetic_Dpi}
\end{table}
The errors in Table~\ref{tab:synthetic_Dpi} take into account the uncertainties induced by:
\begin{itemize}
\item the statistical noise and the fitting procedure;
\item the errors in the determinations of the input parameters, namely the values of the average $u/d$ quark mass $m_{ud}$ the value of the charm quark mass $m_c$, the lattice spacing $a$ and the $SU(2)$ ChPT LECs $f$ and $B_0$, determined in Ref.~\cite{Carrasco:2014cwa}; 
\item the chiral extrapolation, evaluated combining the results obtained using the SU(2) ChPT fit on all our lattice data and a fit with $b_2 = 0$ in Eq.~(\ref{eq:ChLim}) applied only to the data with $M_\pi < 390$ MeV\footnote{In this case the total number of data values reduces to $814$, since the results for the gauge ensembles A80.24, A100.24, B85.24 and B75.32 (see Table~\ref{tab:simu&masses}) are excluded from the analysis.}; 
\item the FSEs, evaluated by comparing the results obtained with and without the FSE factor (\ref{eq:FSE});
\item the discretization effects, calculated by comparing our main results with those obtained including in Eqs.~(\ref{eq:z-exp_f+}-\ref{eq:z-exp_f0}) extra terms proportional to $(a \Lambda_{\rm{QCD}})^4$. 
Using a value for $\Lambda_{QCD}$ equal to $\simeq 0.35$ GeV, we expect that the values of the coefficients of the extra terms are in a natural range of order ${\cal{O}}(1)$. Therefore, we adopt for the coefficients of the extra terms a (conservative) prior distribution equal to $0 \pm 3$.
\end{itemize}

Similarly, in Table~\ref{tab:synthetic_DK} we provide a set of synthetic data points for the vector and scalar $D \to K$ form factors, $f^{DK}_+(q^2)$ and $f^{DK}_0(q^2)$, with the corresponding total uncertainties for eight selected values of $q^2$ between $0$ and $q^2_{\rm{max}} = (M_D - M_K)^2$.
Note that, at variance with the case of the $D \to \pi$ transition, the uncertainty related to FSEs is not considered, because the data for the $D \to K$ transition do not show any visible FSE. 
\begin{table}[htb!]
\renewcommand{\arraystretch}{1.2} 
\begin{center}
\begin{tabular}{|c|c|c|}
\hline
$q^2 ~ (\mbox{GeV}^2)$ & $f_+(q^2)$ & $f_0(q^2)$  \\ \hline 
 $0.0$    & ~ $0.765 ~ (29) ~ (11) ~ (1) ~ [31]$ ~ & ~ $0.765 ~ (29) ~ (11) ~ (1) ~ [31]$ ~ \\ 
 $0.2692$ & ~ $0.815 ~ (28) ~ (12) ~ (1) ~ [31]$ ~ & ~ $0.792 ~ (26) ~ (10) ~ (1) ~ [28]$ ~  \\  
 $0.5385$ & ~ $0.872 ~ (28) ~ (13) ~ (1) ~ [31]$ ~ & ~ $0.820 ~ (23) ~ (10) ~ (1) ~ [25]$ ~  \\  
 $0.8077$ & ~ $0.937 ~ (28) ~ (15) ~ (1) ~ [32]$ ~ & ~ $0.849 ~ (21) ~~ (9) ~ (1) ~ [23]$ ~  \\  
 $1.0769$ & ~ $1.013 ~ (29) ~ (17) ~ (1) ~ [34]$ ~ & ~ $0.879 ~ (19) ~~ (9) ~ (1) ~ [21]$ ~  \\  
 $1.3461$ & ~ $1.102 ~ (32) ~ (21) ~ (1) ~ [38]$ ~ & ~ $0.911 ~ (17) ~~ (9) ~ (1) ~ [19]$ ~  \\  
 $1.6154$ & ~ $1.208 ~ (36) ~ (26) ~ (1) ~ [44]$ ~ & ~ $0.944 ~ (17) ~~ (8) ~ (1) ~ [19]$ ~  \\  
 $1.8846$ & ~ $1.336 ~ (43) ~ (32) ~ (1) ~ [54]$ ~ & ~ $0.979 ~ (17) ~~ (8) ~ (1) ~ [19]$ ~  \\ \hline
\end{tabular}
\end{center}
\renewcommand{\arraystretch}{1.0} 
\caption{\it Synthetic data points for the $D \to K$ transition representing our results for the vector and scalar form factors extrapolated to the physical pion point and in the continuum and infinite volume limits for eight selected values of $q^2$ in the range between $q^2 = 0$ and $q^2 = q_{max}^2 = (M_D - M_K)^2 \simeq 1.88 ~ \gev^2$. The errors correspond to the uncertainties related to (statistical + fitting procedure + input parameters), chiral extrapolation and discretization effects, respectively (see text). The errors in squared brackets correspond to the combination in quadrature of the statistical and all systematic errors.  }
\label{tab:synthetic_DK}
\end{table}

In order to allow a direct use of the synthetic data points without using our bootstrap samples, we have calculated the covariance matrix among the synthetic data points contained either in Table~\ref{tab:synthetic_Dpi} or in Table~\ref{tab:synthetic_DK}. 
The corresponding covariance matrices are available upon request to allow to fit our synthetic data with any functional form, that can be adopted for describing the momentum dependence of the form factors.

For a direct use of our results for the form factors $f_{+, 0}^{D \to \pi(K)}$ we provide in the Appendix the values of the parameters of the z-expansions of our global fit after the extrapolations to the physical pion point and to the continuum and infinite volume limits, including the corresponding covariance matrices.

From Tables \ref{tab:synthetic_Dpi} and \ref{tab:synthetic_DK} our results at zero 4-momentum transfer are
 \be
      f_+^{D \to \pi}(0) = 0.612 ~ (35) \qquad ~ , ~ \qquad f_+^{D \to K}(0) = 0.765  ~ (31) ~ ,
 \ee      
which are consistent within the errors with the FLAG~\cite{Aoki:2016frl} averages $f_+^{D \to \pi}(0) = 0.666 ~ (29)$, based on the result of Ref.~\cite{Na:2011mc}, and $f_+^{D \to K}(0) = 0.747 ~ (19)$ from Ref.~\cite{Na:2010uf}. 

Using the experimental values 
 \bea
     |V_{cd}| f_+^{D \pi}(0) = 0.1426 ~ (19) \quad , \qquad   |V_{cs}| f_+^{D K}(0)  = 0.7226 ~ (34)  ~ ,
     \label{eq:resultsHFAG}
 \eea
given by HFAG in Ref.~\cite{Amhis:2016xyh}, we obtain our results for the CKM matrix elements $|V_{cd}|$ and $|V_{cs}|$:
 \bea
     \label{eq:resultsCKM}
     |V_{cd}| & = & 0.2330 ~ (133)_{\rm lat} ~ (31)_{\rm exp} =  0.2330 ~ (137) ~ , \\
     |V_{cs}| & = & 0.945 ~ (38)_{\rm lat} ~ (4)_{\rm exp} = 0.945 ~ (38) ~ ,
 \eea
where the errors are from the lattice calculation and from the experiments respectively, showing that the dominant error is the theoretical one.
Our results (\ref{eq:resultsCKM}) can be compared with the determinations of $|V_{cd}|$ and $|V_{cs}|$ based on the $D$ and $D_s$ leptonic decay constants, $f_D = 207.4 (3.8)$ MeV and $f_{D_s} = 247.2 (4.1)$ MeV, obtained in Ref.~\cite{Carrasco:2014poa} using the same ETMC gauge configurations. 
Using the experimental values of $f_D |V_{cd}| = 46.06 (1.11)$ MeV and $f_{D_s} |V_{cs}| = 250.66 (4.48)$ MeV from Ref.~\cite{Rosner:2015wva} one gets
 \bea
     |V_{cd}| & = & 0.2221 ~ (41)_{\rm lat} ~ (54)_{\rm exp} =  0.2221 ~ (68) ~ , \\
     |V_{cs}| & = & 1.014 ~ (17)_{\rm lat} ~ (18)_{\rm exp} = 1.014 ~ (25) ~ ,
     \label{eq:resultsCKM_allf}
 \eea
where again the errors are from the lattice calculation and from the experiments, respectively. 
At variance with the semileptonic case, the theoretical uncertainties of $|V_{cd}|$ and $|V_{cs}|$, obtained from the leptonic decays, are comparable to (or even smaller than) the experimental ones.

An alternative way to extract the CKM matrix elements $|V_{cd}|$ and $|V_{cs}|$ is to combine directly the momentum dependence of the semileptonic form factors obtained from lattice QCD simulations with the experimental $q^2$-bins of the differential $D \to \pi(K) \ell \nu_\ell$ decay rates.
The application of such a strategy using the form factors determined in this work will be presented elsewhere \cite{Riggio:2017zwh}.
 
\begin{figure}[htb!]
\centering{
\includegraphics[scale=0.60]{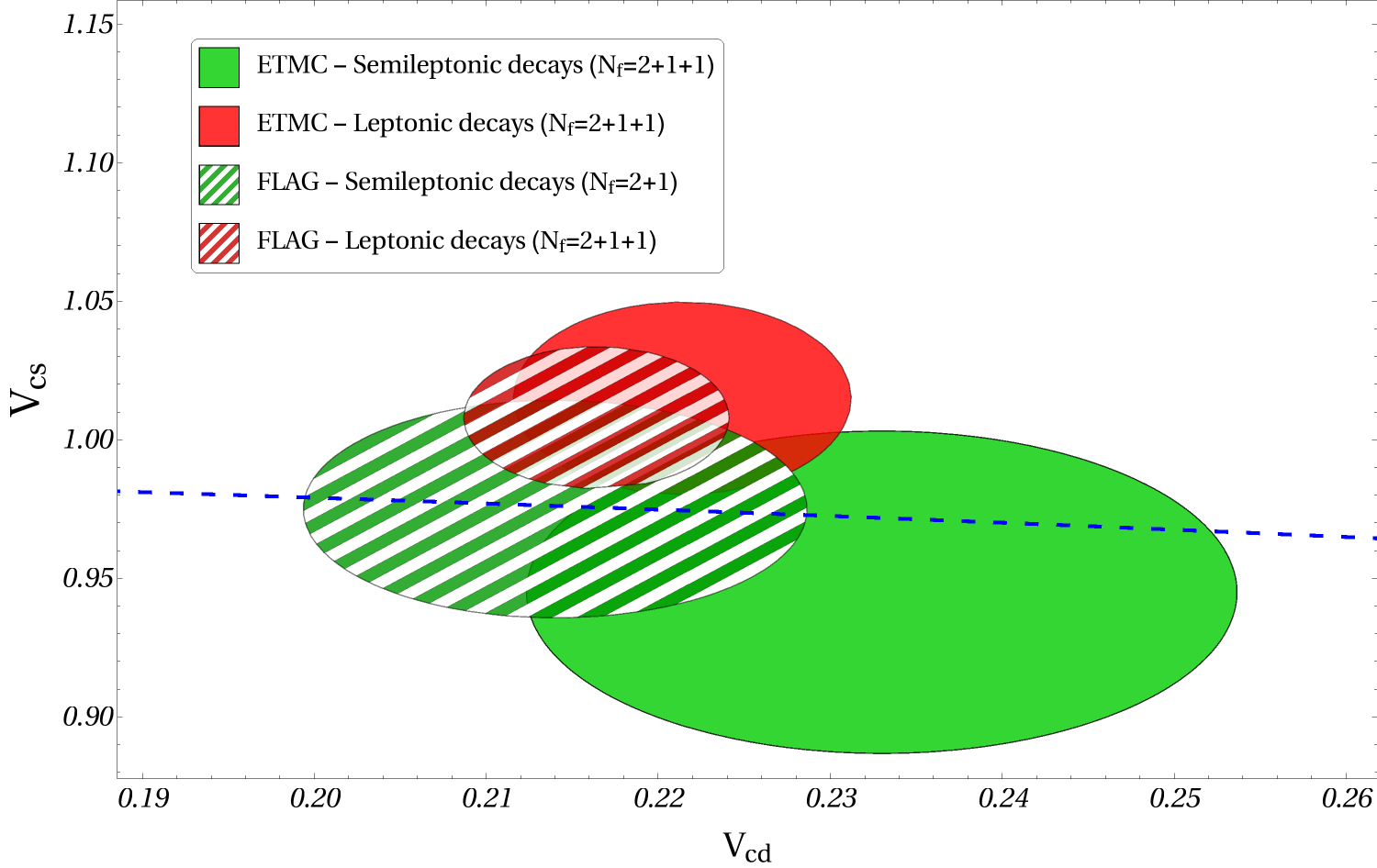}
}
\caption{\it Results for $|V_{cd}|$ and $|V_{cs}|$ obtained from leptonic and semileptonic $D$- and $D_s$-meson decays, represented respectively by green and red ellipses corresponding to a $68 \%$ probability contour. The solid ellipses are the results of Ref.~\cite{Carrasco:2014poa} and of this work, obtained with $N_f = 2 + 1 + 1$ dynamical quarks. The striped ellipses correspond to the latest FLAG results~\cite{Aoki:2016frl}, which for the semileptonic decays are based on the LQCD results obtained in Refs.~\cite{Na:2011mc,Na:2010uf} with $N_f = 2 + 1$ dynamical quarks. The dashed line indicates the correlation between $|V_{cd}|$ and $|V_{cs}|$ that follows if the CKM-matrix is unitary.}
\label{fig:VcdVcs}
\end{figure}

The results (\ref{eq:resultsCKM}-\ref{eq:resultsCKM_allf}) are presented in Fig.~\ref{fig:VcdVcs} as ellipses in the $|V_{cd}| - |V_{cs}|$ plane corresponding to a $68 \%$ probability contour. 
In Fig.~\ref{fig:VcdVcs} also the ellipses corresponding to the leptonic and semileptonic FLAG averages~\cite{Aoki:2016frl} for $|V_{cd}|$ and $|V_{cs}|$ are shown, as well as the constraint imposed by the second-row unitarity, indicated by a dashed line. 
 
Using $|V_{cb}| = 0.0360 (9)$ from Ref.~\cite{Olive:2016xmw} we can perform the check of the unitarity of the second row of the CKM matrix. 
We find
 \bea
        \label{eq:utest}
        |V_{cd}|^2 + |V_{cs}|^2 + |V_{cb}|^2 & = & 0.949 ~ (78) ~ \qquad \mbox{from semileptonic decays} ~ , \nonumber \\[2mm]
        |V_{cd}|^2 + |V_{cs}|^2 + |V_{cb}|^2 & = & 1.079 ~ (54) ~ \qquad \mbox{from leptonic decays} ~ ,
 \eea
which test the second-row unitarity at the level of several percent for both semileptonic and leptonic modes.

\section{Conclusions}
\label{sec:conclusions}

We have presented the first lattice $N_f = 2 + 1 + 1$ determination of the vector and scalar form factors of the $D \to \pi \ell \nu$ and $D \to K \ell \nu$ semileptonic decays, which are relevant for the extraction of the CKM matrix elements $\lvert V_{cd} \rvert$ and $\lvert V_{cs} \rvert$ from experimental data. 

Our analysis is based on the gauge configurations produced by ETMC with $N_f = 2 + 1 + 1$ flavors of dynamical quarks at three different values of the lattice spacing with pion masses as small as 210 MeV. 
Quark momenta are injected on the lattice using non-periodic boundary conditions. 
The matrix elements of both vector and scalar currents are determined for a plenty of kinematical conditions in which parent and child mesons are either moving or at rest. 

Lorentz symmetry breaking due to hypercubic effects is clearly observed in the data and included in the decomposition of the current matrix elements in terms of additional form factors.
We found evidence that the hypercubic artifacts may be governed by the difference between the parent and the child meson masses.
This represents a very important issue, which warrants further investigations, since it might become particularly relevant in the case of the determination of the form factors governing semileptonic $B$-meson decays into lighter mesons.

In the present work the values of the quark momentum were not chosen having in mind the investigation of hypercubic effects in the semileptonic form factors. 
In particular the use of spatially symmetric values of the quark momentum is not ideal for such a purpose. 
We have planned to perform new calculations of the semileptonic form factors removing the above constraint and optimizing the choice of the non-periodic boundary conditions. Nevertheless, we stress that the main structure of the hypercubic effects on the matrix elements of the vector and scalar currents has been understood in the present work.

After the extrapolations to the physical pion mass and to the continuum limit we determine the vector and scalar form factors in the whole kinematical region from $q^2 = 0$ up to $q^2_{\rm max} = (M_D - M_{\pi(K)})^2$ accessible in the experiments, obtaining a good overall agreement with experiments.
Some deviations are visible at high values of $q^2$ for both $D \to \pi \ell \nu$ and $D \to K \ell \nu$ decays.
A set of synthetic data points, representing our results for $f_+^{D \pi(K)}(q^2)$ and $f_0^{D \pi(K)}(q^2)$ for several selected values of $q^2$, is provided, including also the covariance matrix for the data at different values of $q^2$.

At zero 4-momentum transfer we get
 \bea
      f_+^{D \to \pi}(0) & = & 0.612 ~ (35) ~ , \\[2mm]
      f_+^{D \to K}(0) & = & 0.765 ~ (31) ~ . 
 \eea
Using the experimental values (\ref{eq:resultsHFAG}) for $\lvert V_{cd}\rvert f_+^{D \to \pi}(0)$ and $\lvert V_{cs}\rvert f_+^{D \to K}(0)$ from HFAG~\cite{Amhis:2016xyh} we determine
 \bea
     \lvert V_{cd} \rvert & = & 0.2330 ~ (137) ~ , \\[2mm]
     \lvert V_{cs} \rvert & = & 0.945 ~ (38) ~ .
 \eea
Including also the determination of $|V_{cb}|$ from $B$-meson decays \cite{Olive:2016xmw}, the test of the second-row of the CKM matrix is
 \be
      |V_{cd}|^2 + |V_{cs}|^2 + |V_{cb}|^2 = 0.949 ~ (78) ~ .
 \ee

\section*{Acknowledgements}
We warmly thank N.~Carrasco and F.~Sanfilippo for their valuable contribution to the initial stages of the present work.
We thank our colleagues of the ETMC for fruitful discussions.
We gratefully acknowledge the CPU time provided by PRACE under the project PRA067 {\it ``First Lattice QCD study of B-physics with four flavors of dynamical quarks"} and by CINECA under the specific initiative INFN-LQCD123 on the BG/Q system Fermi at CINECA (Italy).
V.L., S.S. and C.T.~thank MIUR (Italy) for partial support under the contract PRIN 2010-2011 and PRIN 2015.

\appendix
\section*{Appendix: The z-expansion of the physical vector and scalar form factors}

After the extrapolations to the physical pion point and to the continuum and infinite volume limits, the z-expansions of the vector and scalar form factors, adopted in this work, are written in the case of the $D \to \pi$ transition as
 \bea
   \label{eq:DPi_f+}
    f_+^{D \to \pi}(q^2) & = & \frac{f^{D \to \pi}(0) + c_+^{D \to \pi} (z - z_0)
                                             \left(1 + \frac{z + z_0}{2} \right)}{1 - P_V ~ q^2} ~ ,\\
   \label{eq:DPi_f0}
    f_0^{D \to \pi}(q^2) & = & \frac{f^{D \to \pi}(0) + c_0^{D \to \pi} (z - z_0) 
                                             \left(1 + \frac{z + z_0}{2} \right)}{1 - P_S ~ q^2} ~ .
 \eea
The values of the five parameters $f^{D \to \pi}(0)$, $c_+^{D \to \pi}$, $P_V$, $c_0^{D \to \pi}$ and $P_S$ are collected in Table~\ref{tab:Dpi_parms}, with the corresponding covariance matrix given in Table~\ref{tab:Dpi_cov}.

\begin{table}[htb!]
\renewcommand{\arraystretch}{1.2} 
\begin{center}
\begin{tabular}{|c|c|c|c|c|}
\hline
$f^{D \to \pi}(0)$ & $c_+^{D \to \pi}$ & $P_V~(\mbox{GeV}^{-2})$  & $c_0^{D \to \pi}$ & $P_S~(\mbox{GeV}^{-2})$ \\ \hline 
$0.6117 ~ (354)$ & $-1.985 ~ (347)$ & $0.1314 ~ (127)$ & $-1.188 ~ (256)$ & $0.0342 ~ (122)$ \\
\hline
\end{tabular}
\end{center}
\renewcommand{\arraystretch}{1.0} 
\caption{\it Values of the parameters appearing in the z-expansions of the vector and scalar form factors (\ref{eq:DPi_f+}-\ref{eq:DPi_f0}) in the case of the $D \to \pi$ transition.}
\label{tab:Dpi_parms}
\end{table}

\begin{table}[h!]
\small
\renewcommand{\arraystretch}{1.2} 
\begin{center}
\begin{tabular}{|c||c|c|c|c|c|}
\hline
& $f^{D \to \pi}(0)$ & $c_0^{D \to \pi}$ & $c_+^{D \to \pi}$ & $P_S$  & $P_V$ \\ \hline 
$f^{D \to \pi}(0)$ & $1.25642 \cdot 10^{-3}$ & $7.18296 \cdot 10^{-3}$ & $6.77051 \cdot 10^{-3}$ & $3.66997 \cdot 10^{-5}$ & $2.87257 \cdot 10^{-5}$ \\ 
$c_0^{D \to \pi}$ & $7.18296 \cdot 10^{-3}$ & $6.56690 \cdot 10^{-2}$ & $6.30124 \cdot 10^{-2}$ & $1.73569 \cdot 10^{-3}$ & $8.43689 \cdot 10^{-4}$ \\ 
$c_+^{D \to \pi}$ & $6.77051 \cdot 10^{-3}$ & $6.30124 \cdot 10^{-2}$ & $1.20371 \cdot 10^{-1}$ & $2.24220 \cdot 10^{-3}$ & $3.25631 \cdot 10^{-3}$ \\ 
                $P_S$ & $3.66997 \cdot 10^{-5}$ & $1.73569 \cdot 10^{-3}$ & $2.24220\cdot 10^{-3}$ & $1.48010 \cdot 10^{-4}$ & $9.60595 \cdot 10^{-5}$ \\ 
                $P_V$ & $2.87257\cdot 10^{-5}$ & $8.43689 \cdot 10^{-4}$ & $3.25631 \cdot 10^{-3}$ & $9.60595 \cdot 10^{-5}$ & $1.60179 \cdot 10^{-4}$ \\ 
\hline
\end{tabular}
\end{center}
\renewcommand{\arraystretch}{1.0} 
\caption{\it Covariance matrix corresponding to the z-expansions of the vector and scalar form factors (A1)-(A2) in the case of the $D \to \pi$ transition.}
\label{tab:Dpi_cov}
\end{table}

Analogously in the case of the $D \to K$ transition the z-expansions of the vector and scalar form factors read as 
 \bea
   \label{eq:DK_f+}
    f_+^{D \to K}(q^2) & = & \frac{f^{D \to K}(0) + c_+^{D \to K} (z - z_0)
                                           \left(1 + \frac{z + z_0}{2} \right)}{1 - q^2 / M_{D_s^*}^2} ~ ,\\
   \label{eq:DK_f0}
    f_0^{D \to K}(q^2) & = & f^{D \to K}(0) + c_0^{D \to K} (z - z_0) \left(1 + \frac{z + z_0}{2} \right) ~ ,
 \eea
where the values of the three parameters $f^{D \to K}(0)$, $c_+^{D \to K}$ and $c_0^{D \to K}$ are collected in Table~\ref{tab:DK_parms}, with the corresponding covariance matrix given in Table~\ref{tab:DK_cov}.

\begin{table}[htb!]
\renewcommand{\arraystretch}{1.2} 
\begin{center}
\begin{tabular}{|c|c|c|}
\hline
$f^{D \to K}(0)$ & $c_+^{D \to K}$ & $c_0^{D \to K}$ \\ \hline 
$0.7647 ~ (308)$ & $-0.066 ~ (333)$ & $-2.084 ~ (283)$ \\
\hline
\end{tabular}
\end{center}
\renewcommand{\arraystretch}{1.0} 
\caption{\it Values of the parameters appearing in the z-expansions of the vector and scalar form factors (\ref{eq:DK_f+}-\ref{eq:DK_f0}) in the case of the $D \to K$ transition.}
\label{tab:DK_parms}
\end{table}

\begin{table}[htb!]
\renewcommand{\arraystretch}{1.2} 
\begin{center}
\begin{tabular}{|c||c|c|c|}
\hline
& $f^{D \to K}(0)$ & $c_0^{D \to K}$ & $c_+^{D \to K}$ \\ \hline 
$f^{D \to K}(0)$ & $9.50493 \cdot 10^{-4}$ & $6.92027 \cdot 10^{-3}$ & $5.66397 \cdot 10^{-3}$ \\ 
$c_0^{D \to K}$ & $6.92027 \cdot 10^{-3}$ & $7.99358 \cdot 10^{-2}$ & $7.55735 \cdot 10^{-2}$ \\ 
$c_+^{D \to K}$ & $5.66397 \cdot 10^{-3}$ & $7.55735 \cdot 10^{-2}$ & $1.10925 \cdot 10^{-1}$ \\ 
 \hline
\end{tabular}
\end{center}
\renewcommand{\arraystretch}{1.0} 
\caption{\it Covariance matrix corresponding to the z-expansions of the vector and scalar form factors (\ref{eq:DK_f+}-\ref{eq:DK_f0}) in the case of the $D \to K$ transition.}
\label{tab:DK_cov}
\end{table}

\bibliographystyle{JHEP}

\bibliography{rifbiblio}

\end{document}